\documentclass[prd,twocolumn,floats,superscriptaddress,eqsecnum,floatfix,
showpacs,
]{revtex4-1}

\usepackage{amssymb,amsmath}
\usepackage{epsfig}

\pdfoutput=1
\usepackage{amssymb,amsmath}
\usepackage{epsfig}
\usepackage{color}
\usepackage{datetime}
\usepackage[utf8]{inputenc}

\newcommand{\pa}{\partial}

\newcommand{\be}{\begin{equation}}
\newcommand{\ee}{\end{equation}}
\newcommand{\ba}{\begin{eqnarray}}
\newcommand{\ea}{\end{eqnarray}}
\newcommand{\beg}{\begin{gather*}}
\newcommand{\eng}{\end{gather*}}

\newcommand{\hh}{,\hspace{0.5cm}}
\newcommand{\hhh}{,\hspace{0.2cm}}

\newcommand{\eq}[1]{(\ref{#1})}

\newcommand{\n}[1]{\label{#1}}

\newcommand{\hor}{\stackrel{ {\mbox{\tiny H}}}{=} }

\newcommand{\ins}[1]{{\mbox{\tiny #1}}}

\newcommand{\const}{\mbox{const}}

\def\XXint#1#2#3{{\setbox0=\hbox{$#1{#2#3}{\int}$ }
\vcenter{\hbox{$#2#3$ }}\kern-.6\wd0}}

\begin{document}

\title{Quantum radiation from a sandwich  black hole}

\author{Valeri P. Frolov}
\email{vfrolov@ualberta.ca}
\affiliation{Theoretical Physics Institute, Department of Physics\\
University of Alberta, Edmonton, Alberta, Canada T6G 2E1}

\author{Andrei Zelnikov}
\email{zelnikov@ualberta.ca}
\affiliation{Theoretical Physics Institute, Department of Physics\\
University of Alberta, Edmonton, Alberta, Canada T6G 2E1}

\begin{abstract}
We discuss quantum radiation of a massless scalar field from a spherically symmetric nonsingular black hole with finite lifetime. Namely, we discuss a sandwich black-hole model, where a black hole is originally created by a collapse of a null shell of mass $M$, and later, after some time $\Delta V$, it is disrupted by the collapse of the other shell with negative mass $-M$. We assume that between the shells the metric is static and either coincides with the Hayward metric or with a special generalization of it. We show that in both cases for sufficiently large parameter $\Delta V$ the radiation after the formation of the black hole practically coincides with the Hawking result. We also calculated the radiation, emitted from the black hole interior. This radiation contains peak at the moment when the second shell intersects the inner horizon. In the standard sandwich metric (with the Hayward interior) this outburst of the energy is exponentially large. In the modified metric, which includes additional non-trivial redshift parameter, this exponent is suppressed. This is a result of significant decrease of the surface gravity of the inner horizon in the latter case. We discuss possible consequences of this result in context of the self-consistency requirement for nonsingular models with quantum radiation.


\end{abstract}


\maketitle


\section{Introduction}

The general relativity is a remarkable theory which allows one to understand spacetime and matter properties in a strong gravitational field. It predicts black holes and is important for description of the early Universe. The general relativity predictions in the weak field approximation are confirmed by observations. Recent discovery by LIGO of the coalescence of two black holes indicated that the general relativity is valid in the strong field regime, when its non-linear effects are important. At the same time the Einstein general relativity theory is ultraviolet (UV) incomplete. Its well known problem is existence of singularities. Solutions of the Einstein equations, describing stationary black holes, have curvature singularities in their interior. It is generally believed that the theory should be modified in the domain where the curvature becomes high. There were proposed several different approaches how to deal with this problem. Theories of quantum gravity, such as the string theory and the loop gravity, are well known examples. Recently a new, very promising UV-complete modifications of the general relativity was proposed \cite{Tomboulis:1997gg, Biswas:2011ar, Modesto:2011kw, Modesto:2012ys, Biswas:2013cha, Biswas:2013kla,
Modesto:2014lga, Talaganis:2014ida,Tomboulis:2015gfa,Tomboulis:2015esa}. It is called a ghost-free gravity. Such a theory contains an infinite number of derivatives and, in fact, is non-local. Similar theory   appears naturally also in the context of noncommutative geometry deformation of the Einstein gravity \cite{Nicolini:2005vd,Spallucci:2006zj} (see a review \cite{Nicolini:2005zi} and references therein).  The application of the ghost free theory of gravity to the problem of singularities in cosmology and black holes can be found in
\cite{Biswas:2010zk, Modesto:2010uh, Hossenfelder:2009fc, Calcagni:2013vra,
Zhang:2014bea, Conroy:2015wfa,Frolov:2015bia,Frolov:2015bta, Frolov:2015usa,Li:2015bqa,Bambi:2016uda,Bambi:2016yne,Bambi:2016wmo,Bambi:2016wdn}. In spite of a number of promising results, we still do not have a final conclusive solution of this problem in the ghost-free gravity. The reason is complexity of its equations, which are both non-local and non-linear.

There exists another approach, which became quite popular recently. Roughly speaking it can be formulated as follows. Suppose there exists such a fundamental theory of gravity, in which black-hole solutions are regular and the singularities are absent. What might be properties of black holes in such a theory. It is natural to expect, that the modified theory of gravity should include some fundamental length scale $\ell$ (or a related mass-scale parameter $\mu=\ell^{-1}$), and the Einstein gravity gives an accurate description of the spacetime geometry in the domain where the curvature is less than $\ell^{-2}$. There were proposed large number of nonsingular models of black holes (see e.g. discussion and references in \cite{Frolov:2016pav} ).

A natural requirement to nonsingular black-hole metrics is that at large radius they correctly reproduce Schwarzschild, Kerr or other black hole solutions of the general relativity. This means that the corresponding nonsingular black-hole metric possesses one or more arbitrary parameters (such as mass, angular momentum and charge). The curvature inside the black hole, being regular, nevertheless depends on the value of these parameters. In a general case it may infinitely grow for a special limit of the parameters. The requirement, that it does not happen and the curvature always remain finite and limited by some fundamental value ($\sim \ell^{-2}$), can be imposed as an additional principle, which restricts the variety of nonsingular black-hole models. The limiting curvature principle was first formulated by Markov \cite{Markov:1982,Markov:1984ii} (see also \cite{Polchinski:1989ae}). For a spherically symmetric black hole this principle, in particular, implies that the apparent horizon cannot cross the center $r=0$. In other words, besides the outer part of the apparent horizon there should exist its inner part, separated from the center. When a black hole evaporates the event horizon does not exist and the apparent  horizon is closed. Such a model was first proposed in  \cite{Frolov:1981mz}, and later was intensively discussed in the literature \cite{Roman:1983zza,Borde:1996df,Hayward:2005gi,Hossenfelder:2009fc,Bambi:2013gva,Bambi:2013caa, Hawking:2014tga, Frolov:2014jva,Bardeen:2014uaa,Haggard:2014rza, Barrau:2015uca,Haggard:2015iya}.

A general feature of nonsingular models of a black hole with a closed apparent horizon is the following. All the quanta which either fell into the black hole or were created inside it are finally emitted to an external observer after the complete evaporation of the black hole.
Outgoing radial null rays in the black-hole interior are accumulated in the vicinity of the inner horizon. This is a consequence of negative value of the surface gravity at the inner horizon. As a result one can expect that  particles emitted from the inner horizon at the final stage of the black hole evaporation would have large blueshift. In a self-consistent model, when the back-reaction of the created particles is properly taken into account, this final burst of the radiation should be somehow suppressed. In the absence of the theory of gravity which properly describes the black-hole interior and the final state of the evaporating black hole, one can estimate the quantum radiation in the adopted nonsingular black hole model. In such a case a consistency requirement can be used as an additional test of the plausibility of the model \cite{Bolashenko:1986mr}.

The present paper discusses this problem. Namely, we study quantum radiation of massless particles from nonsingular black holes. To attack this problem we assume a number of simplifications. To describe a spherically symmetric black hole which has finite life-time we consider the following model. We assume that a black hole is formed a result of the collapse of the null shell of positive mass $M$ and ends its existence as a result of collapse of another null shell with negative mass $-M$. We call it a {\it sandwich black hole}. Such a model was also considered earlier in the interesting paper \cite{Bianchi:2014bma}, where  the problem of the black hole entropy was discussed.  Certainly such a model is quite different from a ``real'' evaporating black hole. However, it happens that because of its simplicity it might be useful for a study of quantum effects in black holes with a closed apparent horizon. We shall argue that some of its predictions might be quite robust and remain valid for  more realistic ``smooth models''. To estimate the quantum radiation of the massless particles we use a
the result by Christensen and Fulling \cite{Christensen:1977jc}, who derived the  two-dimensional quantum average of the stress-energy tensor from the conformal anomaly. It is well known that this approximation gives quite good result for Hawking radiation (see e.g. \cite{Frolov:1998wf} ).

The paper is organized as follows. In section~$II$ we discuss generic properties of nonsingular black holes and collect some useful formulas for the quantum energy flux at the infinity in an asymptotically flat spacetime. We also discuss a classical scattering problem for a massless particle propagating along radial geodesics from the past to future null infinity.
We derive a useful expression for a  gain function, which is the ratio of its final and initial energies. In section~$III$ we describe a sandwich model of a nonsingular black hole. This section also contains general expressions for the gain function and quantum energy flux for a sandwich nonsingular black hole. In section~$IV$ we discuss
a special case, when the metric between the null shells coincides with the Hayward metric \cite{Hayward:2005gi}. Such a ``standard'' model is characterised by two parameters, its mass and time duration of its existence. We present both analytic and numerical results and discuss the behavior of the quantum radiation during the time of the black-hole formation, existence, and its disruption. In section~$V$ we discuss a sandwich model where the regular metric between the shells contains a nontrivial redshift factor. We calculate the quantum energy flux from such a modified sandwich black hole and demonstrate that the exponentially large peak of the radiation from the inner horizon, which is  present in the ``standard'' model, can be suppressed by a proper choice of the redshift function. This suppression effect is a consequence of reduction of the (negative) surface gravity of the inner horizon. Discussion of the obtained results and additional remarks are presented in section~$VI$.

\section{Non-singular models of evaporating black holes}

\subsection{Spherically symmetric regular black holes}

The most general spherically symmetric metric in the four dimensional spacetime can be written in the form
\be\n{a.1}
dS^2=-A^2 F dV^2+2A dV dR+R^2 d\omega^2\, .
\ee
It contains two arbitrary functions of the advanced time $V$ and radius $R$. These coordinates have the dimensionality of $[ length]$, and the dimensionality of $dS^2$ is $[ length^2]$. In what follows it is convenient to deal with dimensionless objects. For this purpose one can use any length-parameter as a standard scale. For example, in the metrics that we shall consider later it might be some fundamental scale $\ell$, or the gravitational radius of the black hole. Sometimes, it is convenient to use their combination, or other scales. We denote this, unspecified at the moment, standard length-scale by $\sigma$ and denote
\ba\n{sig}
&&dS^2=\sigma^2 ds^2\hhh V=\sigma v\hhh R=\sigma r\, ,\nonumber\\
&&F(V,R)=f(v,r)\hhh A (V,R)=\alpha(v,r)\, .
\ea
Thus one has
\be \n{a.0}
 ds^2=-\alpha^2 f dv^2 +2\alpha dv dr +r^2 d\omega^2\, .
\ee

It is easy to check that
\be
 f=g^{rr}=g^{\mu\nu}\nabla_{\mu}r\nabla_{\nu}r\, .
\ee
Points where $f=0$ form an apparent horizon. We assume that the spacetime is asymptotically flat. In this case the function $f$ at spatial infinity must take the value 1 in order to escape a solid angle deficit
\be
f(v,r)|_{r\to\infty}=1\, .
\ee
Using an ambiguity in the choice of $v$, we impose the following (gauge fixing) condition
\be\n{aav}
\alpha (v,r)|_{r\to\infty}=1\, .
\ee

In what follows we consider, so called, non-singular black holes (see, e.g.,  \cite{Frolov:2016pav} and references therein). One of the conditions, which is valid for such metrics, is the requirement of the regularity of the metric at the center $r=0$. Let $R$ be the Ricci scalar, $S_{\mu\nu}=R_{\mu\nu}-{1\over 4}g_{\mu\nu}R $, and $C_{\mu\nu\alpha\beta} $ be the Weyl tensor. Let us define the following quadratic in the curvature invariants
\be\n{a.4}
{\cal S}^2=S_{\mu\nu} S^{\mu\nu}\hh
{\cal C}^2=C_{\mu\nu\alpha\beta} C^{\mu\nu\alpha\beta}\, .
\ee
We call a metric \eq{a.1} finite at the center, if the metric functions $f$ and $\alpha $ have the following expansions
\ba
f(v,r)&=&f_0(v)+f_1(v) r+f_2(v) r^2+\ldots\, ,\\
\alpha (v,r)&=&\alpha_0(v)+\alpha_1(v) r+\alpha_2(v) r^2+\ldots \, .
\ea
We call a finite at the center metric \eq{a.1} {\em regular} if the invariants $R$, ${\cal S}^2$ and
${\cal C}^2$ are finite at $r=0$. This regularity condition implies that
\be \n{FAc}
f_0(v)=1\hh f_1(v)=\alpha_1(v)=0\,  .
\ee
An important consequence is that for a non-singular black the apparent horizon cannot cross the center $r=0$. In a general case, when $\alpha(v)\ne 1$, the rate of  the proper time $\tau$ at the center differs from the rate of time $v$
\be
d\tau=\alpha_0(v) dv\, .
\ee
The conditions (\ref{FAc}) imply that the geometry near $r=0$ is locally flat, and, in particular, there is no solid-angle deficit.

\subsection{Static spacetime}\label{Static}

Let us make a few remarks, concerning a special case of the metric (\ref{a.1}), when both metric functions, $f$ and $\alpha$, are time independent. Denote by $\xi=\xi^{\alpha}\pa_{\alpha}=\pa_v$ the corresponding Killing vector. Then one has
\be
\xi^2=-\alpha^2 f\, .
\ee
One can check that for such a metric the Killing horizon coincides with the apparent horizon, and $\alpha$ is regular at the horizon function. The surface gravity ${\kappa}$ is defined as follows
\be
\xi^{\beta} \xi_{\alpha;\beta}\hor{\kappa} \xi_{\alpha}\, .
\ee
Simple calculations give
\be \n{kkk}
{\kappa}={1\over 2}\left.\left( \alpha \pa_r f\right)\right|_{H}\, .
\ee

\subsection{Regular evaporating black hole models}

We assume now that a regular metric \eq{a.0} describes a black hole, which was created as a result of the spherical collapse, and it disappears  after some finite time, for example, as a result of its quantum evaporation. For such a system there exist parameters $v^-$ and $v^+$, such that
\be
f=\alpha=1, \mbox{  for $v<v^-$ and $v>v^+$}\, .
\ee
For definiteness, we can choose $v^-$ and $v^+$ so, that in the domain $v^-<v<v^+$ the spacetime curvature does not vanish. The condition (\ref{aav}) fixes the coordinate $v$ up to a constant. We use this freedom to put $v^-=0$. We also denote $q=v^+$.

Consider an incoming radial null ray described by the equation $v=\mbox{const}$ . It propagates from the past null infinity, ${\cal I}^-$, and  reaches the center $r=0$. After passing the center, it becomes an outgoing radial null ray. We shall use diagrams where the angle variables are suppressed, so that the radial null ray will be presented by a line which is reflected at the origin $r=0$. We choose the retarded null time parameter $u_-$ so that at $r=0$ one has $u_-=v$. In the initial flat domain, where $v<0$, one has $u_-=v-2r$. However, in a general case, for $v>0$ this relation between $u_-$ and $v$ is not valid. In particular, in the final flat domain, where $v>q$,  the null coordinate $u_+=v-2r$ differs from $u_-$, and one has relations
\be
u_+=u_+(u_-)\hh u_-=u_-(u_+)\, .
\ee
One can rewrite the first relation in the form $u_+=u_+(v)$. This relation can be interpreted as a one, establishing a map between ${\cal I}^-$, parameterised by $v$, and ${\cal I}^+$, parameterised by $u_+$.

Let us describe a simple algorithm which allows one for find the required map.

Case I. Consider first outgoing rays with $u_-<0$. They cross both of the null surfaces $v=0$ and $v=q$. The radius of the first intersection is
\be \n{eq1}
r_-=-{1\over 2}u_-\, .
\ee
Denote by $r=r(v)$ a solution on the differential equation
\be \n{eq2}
{dr\over dv}={1\over 2} {\cal F}(v,r)\hh {\cal F}=\alpha f\, ,
\ee
with the initial condition
\be \n{eq3}
r(0)=-u_-/2\, .
\ee

This solution describes an outgoing null ray passing through $(0,r_-)$. Denote by $r_+$ the radius $r$ where this ray crosses the second null shell, $r_+=r(q)$. Since in the final domain, where $v>q$, the spacetime is flat, one has
\be \n{eq4}
u_+=q-2r(q)\, .
\ee
Relations \eq{eq1}--\eq{eq4} determines the required map.

Case II. Let $0<u_-<q$ and let $r(v)$ be a solution of \eq{eq2} with the initial condition $r(u_-)=0$. Then a relation (\ref{eq4}) determines the map.

Case III. Let $u_->q$. Then one has $u_+=u_-$.

\subsection{Quantum fluxes at ${\cal I}^+$}

In what follows, we study quantum radiation of a massless scalar field from regular``black holes''. For the calculation of the Hawking radiation one often uses a decomposition of the quantum modes in spherical harmonics. In such an analysis it was shown that the main contribution to the radiation is given by $S$-modes \cite{Frolov:1998wf}. If one reduces to considering only $S$-modes, the theory is effectively reduced to the quantum theory of the  $2D$ massless scalar field in $(t,r)$ sector of the black hole geometry. We shall use a similar $2D$ approximation for the estimation of the quantum energy fluxes from a regular black hole's \footnote{Certainly, the question of how good is this approximation for the calculation of the energy flux from the black hole interior, should be studied.}

An effective action for a two-dimensional  conformal scalar field is
\be
S=-{1\over 2}\int d^2x\sqrt{-g}\,(\nabla\hat{\varphi})^2,
\ee
where the two-dimensional metric is given by
\be
ds^2=-\alpha^2f dv^2+2\alpha dv dr\, .
\ee
Let us notice that the rate of the energy emission, $\dot{E}$, is a dimensionless quantity. For the quantum radiation it is proportional to the Planck length square, $\ell_{Pl}^2$. Thus one has
\be
\dot{E}=(\ell_{Pl}/\sigma)^2 {\cal E}\, ,
\ee
where ${\cal E}$ is the dimensionless rate of the energy emission.
The energy rate flux of massless particles, created from the initial vacuum state, is given by the following expression
\be\n{EEEE}
{\cal E}={1\over 48\pi}\left[ -2{d^2 P\over du^2}+\left({d P\over du}\right)^2\right]\, ,
\ee
where
\be\label{P}
P=\ln \left| {du_-\over du_+}\right|\, .
\ee
This relation directly follows from a general result obtained by Fulling and Christensen \cite{Christensen:1977jc} for the quantum average of the stress-energy tensor of a massless scalar field in two dimensions, reconstructed from the conformal anomaly. The same expression can be also obtained by the variation of the Polyakov effective action with respect to the metric \cite{Frolov:1984,Vilkovisky:1985}.

The expression (\ref{EEEE}) for the energy flux contains two terms. The second one is a square of the first derivative of $P$ and hence it is always positive. The first one, proportional to the second derivative of $P$, can be both positive and negative. Hence, for some periods of the retarded time $u_+$ the flux of the energy from the black hole can be negative. However, the total  emitted energy is always positive. This can be easily checked. Really
\be
E_{tot}={1\over 48\pi}\left[\int_{-\infty}^{\infty} du_+ \left({d P\over du_+}\right)^2 -
2\left. {d P\over du_+}\right|_{-\infty}^{\infty} \right] \, .
\ee
For a nonsingular black hole, which exits during a finite interval of time, the boundary terms vanish.

Quantum effects in nonsingular black holes were discussed earlier \cite{Bolashenko:1986mr,Lorenzo}.
The main goal of this paper is to study quantum energy flux from a sandwich black hole.  In order to calculate this flux in the adopted $2D$ approximation it is sufficient to study propagation of the radial null geodesics from ${\cal I}^-$ to ${\cal I}^+$. The function $u_-(u_+)$, which establishes a map ${\cal I}^+ \to {\cal I}^-$, allows one to find  the function $P$, which enters the expression (\ref{EEEE}) for the energy flux  on ${\cal I}^+$. In the next section we demonstrate that the function $P$ is a logarithm of the ratio of the energy of a classical massless particle at ${\cal I}^+$  to its initial energy at ${\cal I}^-$. It is interesting that $P$ is also related to so called {\it radiation entropy} $\Delta S_{rad}(u_+)$ at retarded time $u_+$, defined in \cite{Bianchi:2014bma}. Namely, one has
\be
\Delta S_{rad}(u_+)=-{1\over 12} P\, .
\ee

\subsection{Gain function}

To understand better main features of the quantum particle production by a black hole it is instructive to consider propagation of classical massless particles (photons) in its geometry.
Motion of radial incoming null rays in the metric (\ref{a.0}) is rather simple. For this rays one has $v=$const, and the corresponding four-momentum is $l^{\mu}=(0,\dot{r},0,0)$.
The dot means a derivative with respect to the affine parameter $\lambda$ \footnote{ We use an ambiguity in the choice of $\lambda$ in such a way, that $\dot{x}^{\mu}$ coincides with the four-momentum of the photon.}. The geodesic equation for the null ray is
\be\n{geod}
{D^2 x^{\mu}\over d\lambda^2}\equiv {d^2 x^{\mu}\over d\lambda^2}+\Gamma^{\mu}_{\alpha\beta}{d x^{\alpha}\over d\lambda}{d x^{\beta}\over d\lambda}=0\, .
\ee
For the radial rays this equation is identically satisfied in the $(\theta,\phi)$ sector, while the only non-vanishing component of the Christoffel symbol $\Gamma^r_{\mu\nu}$ in the $(v,r)$ sector is
\be
\Gamma^r_{r r}={\pa_r \alpha\over \alpha}\, .
\ee
The equation (\ref{geod}) is identically satisfied for $\mu=v$, and for $\mu=r$ gives
\be
(\alpha \dot{r})\dot{}=0\, .
\ee
This means that the quantity $\alpha \dot{r}$ is constant along the radial incoming null ray. Consider a photon, which starts its motion at ${\cal I}^-$, where $\alpha=1$ and its energy is $E_-=-\dot{r}$.  When such a ray arrives to the center $r=0$, where $f=1$, one has
\be
\dot{r}_0=-E_-/\alpha_0\, .
\ee
Here $\alpha_0$ is the value of the redshift function $\alpha$ at the center.
The geometry near the center is regular. Let us introduce Cartesian coordinates $(X,Y,Z)$ in its vicinity, and choose their orientation so that the in-falling photon passes the center $X=Y=Z=0$ moving in the negative direction along $X$ axis. Before it crosses the center one has $r=X$. After this, the photon continues its motion along $X$ direction, however now, for the outgoing photon one has $r=-X$. This means that
\be
\dot{r}_0=E_-/\alpha_0 \, .
\ee
For the outgoing null ray one has
\be\n{forv}
\dot{r}={1\over 2}{\alpha f} \dot{v} \, .
\ee
and $f=1$ at the center. Thus
\be\n{aEm}
(\alpha \dot{v})_0=2{E_-\over \alpha_0}\, .
\ee

The outgoing null rays obey the equation
\be\n{raf}
{dr\over dv}={1\over 2}{\alpha f}\, .
\ee
In a general case the metric functions $\alpha$ and $f$ depend on both $r$ and $v$.
We denote the four-momentum of the radial outgoing null ray by $k^{\mu}$, and denote by $\lambda$ its affine parameter. Thus one has $k^{\mu}=(\dot{v},\dot{r},0,0)$, where a dot denotes a derivative with respect to $\lambda$. The only non-vanishing components of $\Gamma^v_{\mu\nu}$ in the $(v,r)$ sector is
\be
\Gamma^v_{v v}={1\over 2\alpha} \left[
\alpha^2 \pa_r f +2\alpha f \pa_r \alpha+2\pa_v \alpha\right]\, .
\ee
Using (\ref{raf}) and the relation
\be
\dot{\alpha}=\dot{v}\pa_v\alpha +\dot{r}\pa_r\alpha\, ,
\ee
one can write $v$-component of the equation (\ref{geod}) in the form
\be
(\alpha\dot{v})\dot{}=-{1\over 2}\alpha (\alpha f)' \dot{v}^2\, .
\ee
The first integral of this equation is
\be\n{adaf}
\alpha\dot{v}=(\alpha\dot{v})_{v_0} \exp\left[-{1\over 2}\int_{v_0}^v \pa_r(\alpha f) dv\right]\, .
\ee

Suppose there exists a spacetime domain where the metric is static, so that $\pa_v \alpha=\pa_v f=0$. We denote by $\xi=\xi^{\mu}\pa_{\mu}=\pa_v$ the corresponding Killing vector. Then the energy of a photon, which moves in this domain
\be
E=-g_{\mu\nu} \xi^{\mu} k^{\nu}=\alpha^2 f \dot{v}-\alpha \dot{r}={1\over 2} \alpha^2 f \dot{v}\, ,
\ee
is conserved.

Consider a case of an evaporating black hole. If the spacetime  after black-hole evaporation is flat one can use the relation (\ref{adaf}) as follows. For large $v$ the spacetime is flat, so that $f=\alpha=1$ in this domain, and the energy $E_+$ of the outgoing ray is
\be
E_+={1\over 2} \dot{v}_{ {\cal I}^+}\, .
\ee
Denote by $v_0$ the advanced time $v$ when the incoming null ray crosses the center, then using (\ref{adaf}) and (\ref{aEm}) one gets
\be\n{betagen}
\beta={E_+\over E_-}={1\over \alpha_0} \exp\left[-{1\over 2}\int_{v_0}^{\infty} \pa_r(\alpha f) dv\right]\, .
\ee
We call the ratio $\beta$ of the final energy of a photon to its initial energy  a {\em gain function}.

Let us show now that the gain function is related to the function $P$ defined by (\ref{P}). Namely one has
\be\n{PB}
P=\ln \beta \, .
\ee
Consider two nearby incoming radial null rays with parameters $v_0$ and $v_0+\delta v$. They pass the center $r=0$ with the proper time interval difference $\delta \tau=\alpha(v) \delta v$. When the second incoming ray reaches the center, the first one has already passed it. It crosses the world line of the second ray at a point $(v+\delta v,\delta r_0=\delta \tau)$, where $\delta r|_0=\delta \tau /2$. Using equation (\ref{raf}) one can find how this separation $\delta r$ between two nearby outgoing rays, calculated for $v$=const, changes along the outgoing ray. Namely, one has
\be
{d\delta r\over dv}={1\over 2}\pa_r({\alpha f})\delta r\, .
\ee
Integrating this equation one gets
\be
\delta r|_{{\cal I}^+}=\delta r_0 \exp\left[ {1\over 2} \int_{v_0}^{\infty} \pa_r(\alpha f) dv\right]\, .
\ee
Since for a fixed value $v$ $\delta u_+=2\delta r|_{{\cal I}^+}$ and $\delta u_-=\delta v_0$, using the above results one obtain
\be
{du_-\over du_+}={\delta u_-\over \delta u_+}=\beta \, .
\ee
Thus the relation (\ref{PB}) is proved.

The obtained result has a quite simple explanation. Consider a high-frequency wave packet $\sim \exp[i \Phi(u_-)]$ emitted from ${\cal I}^{-}$. Its frequency, as measured by an observer near ${\cal I}^{-}$ is $\omega_-= d\Phi/d{u_-}$. In the adiabatic $2D$ approximation such a packet, when it arrives to ${\cal I}^+$ has a form $\sim  \exp(i \Phi(u_-(u_+))$, and its frequency is
\be
\omega_+=d\Phi/d{u_+}={d\Phi\over d{u_-}} {du_-\over d{u_+}}=\beta\omega_- \, .
\ee
This relation is in the accordance with the definition of the gain function $\beta$.

\section{Quantum radiation from a sandwich black hole}

\subsection{Double-shell model of a nonsingular black hole}

A black hole after its formation becomes a source of quantum radiation. An external observer registers an outgoing flux of Hawking radiation. As a result, the black hole mass decreases and the black hole shrinks. One of the options is that the black hole completely disappears in this process. For a non-singular spherical black hole this means that the apparent horizon is closed, and the event horizon does not exist. Strictly speaking, this object is not a black hole (according to the standard definition), but its long time ``imitation''. For simplicity, we shall use the same name black hole for these objects as well.

\begin{figure}[tbp]
\centering
\includegraphics[width=4cm]{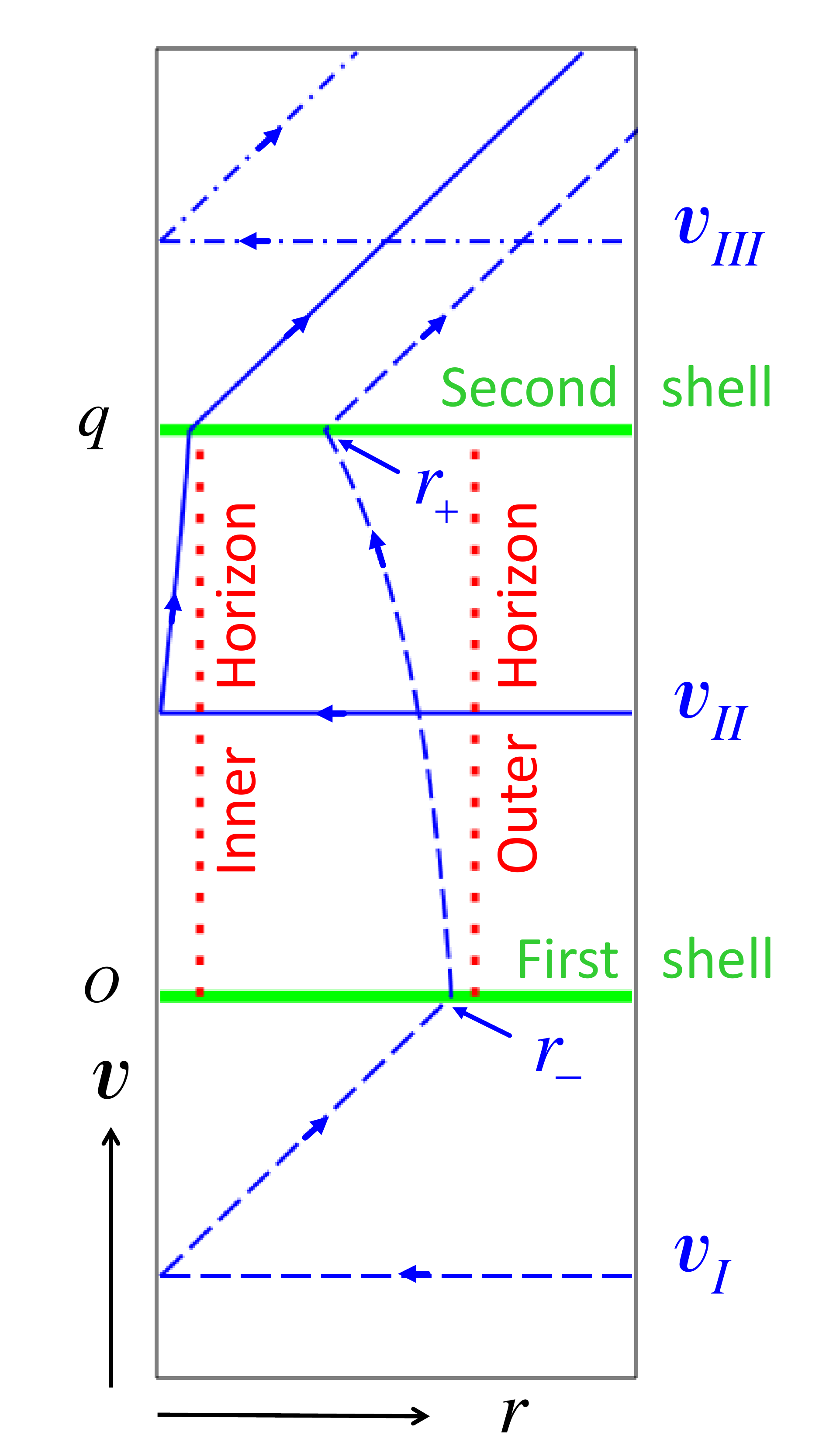}
  \caption{
 Spacetime of  a sandwich black hole in $(v,r)$ coordinates. First null shell with the positive mass is shown by a horizontal line $v=0$, while the second one, with the negative mass, is shown by the horizontal line $v=q$. The spacetime before the first shell and after the second one is flat.  We denote these domains as $I$ and $III$, respectively. The domain $II$ is located between the shells. Inner and outer horizons in the domain $II$ are the inner and outer branches of the apparent horizon, respectively.  Incoming and outgoing radial null rays are schematically shown at this diagram.
\label{Orbits_123}}
\end{figure}

An important new feature of such non-singular black holes is that besides the standard Hawking flux there exists an additional quantum radiation, coming from the black hole interior. One should expect that, if the black hole completely disappears as a result of evaporation, then the total energy loss by the evaporating black hole must be equal to the initial mass of this object. As we shall demonstrate this condition imposes severe self-consistency restriction of the non-singular models of black holes. To illustrate this we consider a simple model. Namely, we assume that a black hole is formed as a result of a spherical collapse of a null shell of mass $M$, it exists for time $\Delta V$, and after disappears, as a result of the collapse of another null shell of mass $-M$ (see Fig.\,\ref{Orbits_123}). During the time interval $\Delta V$ the metric is a static non-singular one. The corresponding metric (in the dimensionless form) is
\be\n{aaff}
ds^2 = \alpha (-{\cal F} dv^2+2 dv dr) +r^2 d\omega^2\, ,
\ee
where ${\cal F}=\alpha f$. $\alpha$ and ${\cal F}$ are functions of $r$ for $0<v<q$, while outside this interval $\alpha={\cal F}=1$. We call such a model a {\it sandwich black hole}. Certainly, this model is quite different from an expected behavior of the evaporating black hole. However, they have the following common feature: a finite time of the `black hole'' existence \footnote{ Similar sandwich type black holes were discussed earlier in the interesting paper \cite{Bianchi:2014bma}, where the problem of the entanglement entropy of evaporating black holes was considered. Namely, the authors of this paper assumed that the metric between two null collapsing shell is constructed by gluing together of the de Sitter metric (for small radius) and the Schwarzschild one (for large radius). They argued that such a metric can be used as an approximation for the Hayward  spacetime. In the first part of our paper we discuss a sandwich black hole with exact Hayward interior. What is more important, we discuss in the second part of the paper metrics between the shells which have an additional new property: a non-trivial redshift factor, which was ignored in the paper \cite{Bianchi:2014bma}.}.

\subsection{Gain function for a sandwich black hole}

For the sandwich black hole with a static interior the formula for the energy gain is simplified. Consider an incoming radial photon with the initial energy $E_i$. After crossing the first shell at the radius $r_-$ it passes through the spacetime between the shells and leaves the sandwich black hole  with the energy $E_f$, crossing the second shell at $r_+$.  One has
\be\label{beta}
\beta={ {\cal F}_-\over {\cal F}_+}={dr_-\over dr_+}\, .
\ee
Here $ {\cal F}_{\pm}$ is a quantity $ {\cal F}$ calculated at points of the entrance, $r_-$, and of the exit, $r_+$, of the photon in the domain between the shells, respectively. In the last equality we used the relation (\ref{betagen}).
For a photon, propagating along a horizon $r_+=r_-$, one has ${\cal F}_-={\cal F}_+=0$. For this case one use the original formula (\ref{adaf}). Since ${1\over 2}(\pa_r(\alpha f)_\ins{H}=\kappa_\ins{H}$, one has
\be
\beta_\ins{H}=\exp(-\kappa_\ins{H} q)\, .
\ee

Consider a beam of radial type $II$ incoming photons with the energy $E_i$ crossing the first shell between $r_-$ and $r_-+\Delta r_-$. They will cross the second shell in the interval $(r_+,r_++\Delta r_+)$ having the energy $E_f$. Then
\be
E_- \Delta r_-=E_+ {dr_-\over dr_+} \Delta r_+ =E_- {{\cal F}_-\over {\cal F}_+} \Delta r_+=E_+ \Delta r_+\, .
\ee
We define the rate of the energy flux ${\cal E}_{\pm}=E_{\pm}/\Delta r_{\pm}$. Then one has
\be
{\cal E}_+=\beta^2 {\cal E}_- \, .
\ee

Let us now calculate the gain function for the type-$II$ photons (see Fig.\,\ref{adaf}). Such a photon starts its motion from ${\cal I}^-$ with $v_0\in (0,q)$. For such an incoming photon $l_{\mu}=(-E_-,0)$, where $E_-$ is its energy at ${\cal I}^-$. After passing $r=0$ this photon becomes outgoing with the same energy, and one has
\be
\ln \beta=-{1\over 2}\int_{v_0}^{q} \pa_r(\alpha f) dv -\ln\alpha_0\, .
\ee
For a double-shell model with a constant metric in the interior one gets
\be
\beta={1\over {\cal F}_+}\, .
\ee

For the type $III$ null ray the gain function is 1.

\subsection{Quantum radiation from a sandwich black hole}

We use dimensionless form of the metric \eq{a.0}, keeping the scale parameter $\sigma$ arbitrary. In the adopted model of a sandwich black hole
\be
f=\alpha=1, \mbox{  for $v<0$ and $v>q$}\, ,
\ee
while for $v\in(0,q)$ these functions  depend only on $r$. The equation (\ref{eq2}) between the shells can be easily solved with the following result. Denote
\be\label{calF}
Q=\int_0^r {dr\over {\cal F} }\hh {\cal F}=\alpha f\, .
\ee
Then the solution is
\be
v-2Q(r)=C=\mbox{const}\, .
\ee

\subsubsection{Type $I$ rays}
This is a case when $u_-<0$, so that the corresponding outgoing null ray intersects both of the null shells. In this case one has
\ba
&&u_+=q-2r_+\hh u_-=-2r_-\, ,\n{uu}\\
&&Q(r_-)=Q(r_+)-q/2\, .\n{QQ}
\ea
Equations (\ref{uu}  establish relations between retarded times $u_+$ and $u_-$ and the radii $r_-$ and $r_+$ of the points where this null ray crosses the null shells, while the  relation (\ref{QQ})  determines a map between $r_-$ and $r_+$. The general expression (\ref{EEEE}) for the energy flux ${\cal E}$ can be transformed in the form more convenient for the calculations.  Using \eq{uu} one gets
\ba
&&P=\ln \left| {dr_-\over dr_+}\right|\, ,\\
&&{\cal E}={1\over 192\pi}\left[ -2{d^2 P\over dr_+^2}+\left({d P\over dr_+}\right)^2\right]\, . \n{EEE}
\ea
Using (\ref{QQ}) one gets
\be\n{FrFr}
{dr_-\over dr_+}={{\cal F}(r_-)\over {\cal F}(r_+)}
\ee
and, hence,
\be\n{Pr}
{d P\over dr_+}={1\over  {\cal F}(r_+)}\Big[{d {\cal F}(r_-)\over dr_-}-{d {\cal F}(r_+)\over dr_+}\Big],
\ee
\be\begin{split}
{d^2 P\over dr_+^2}&={1\over  {\cal F}^2(r_+)}\Big[-{d {\cal F}(r_-)\over dr_-}{d {\cal F}(r_+)\over dr_+}+\Big({d {\cal F}(r_+)\over dr_+}\Big)^2\\
&+
{\cal F}(r_-){d^2 {\cal F}(r_-)\over dr_-^2}-{\cal F}(r_+){d^2 {\cal F}(r_+)\over dr_+^2}\Big].
\n{Prr}
\end{split}\ee

If we introduce the function
\be
B(r)=-2{\cal F}(r){d^2 {\cal F}(r)\over dr^2}+\Big({d{\cal F}(r)\over dr}\Big)^2,
\ee
then
\be\n{bb.2}
{\cal E}={1\over 192\pi}{B(r_-)-B(r_+) \over  {\cal F}^2(r_+) }\hh r_{-}=r_{-}(r_{+})\, .
\ee

\subsubsection{Type $II$ rays}

In this  case  $0<u_-<q$ and the corresponding outgoing null rays intersects only the second  null shell with negative mass. One has
\be
u_+=q-2r_+\hh u_-=q-2Q(r_+)\, .\n{uQQ}
\ee
For the calculation of the energy flux one can use \eq{EEE} with
\be
P=-\ln {\cal F}(r_+)\, .
\ee
To obtain derivatives of $P$ one can use  formulas (\ref{Pr}) and (\ref{Prr}) by putting ${\cal F}(r_-)=1$ in these relations. The final result is
\be\n{bb.22}
{\cal E}=-{1\over 192\pi}{B(r_+) \over  {\cal F}^2(r_+) }\, .
\ee

For the type $III$ rays one has $u_+=u_-$ and the quantum radiation vanishes, ${\cal E}=0$.

\subsection{Sandwich-model with $\alpha=1$}

Let us notice, that in the case when $\alpha=1$ the above expressions for the energy flux can be further simplified. We denote a function $\kappa$ by the relations
\be
\kappa^2=|\xi^2 w^2|\hh w_{\mu}={1\over 2}\nabla_{\mu}\ln|\xi^2|
\hh
|\xi^2|=f\, .
\ee
The quantity $\kappa$ is nothing but the redshifted proper acceleration of the Killing observer at the point $r$
\be
\kappa={1\over 2}{df(r)\over dr}\, .
\ee
Denote by $R(r)$ the scalar curvature at the point $r$
\be
R=-{d^2f(r)\over dr^2}\, .
\ee
Then one can show that the function $B(r)$, which enters relations (\ref{bb.2}) and (\ref{bb.22}), can be written in the form
\be\label{Br}
B(r)=4\kappa^2(r)+2R(r)f(r)\, .
\ee
Both $\kappa$ and $R$ are finite and smooth functions at all radii.

\subsection{Quantum radiation from the near horizon domains}

Let us demonstrate now how the relation (\ref{bb.2}) can be used to calculate exactly the quantum energy flux in some special cases. Namely, we assume that there exists a point $r_\ins{H}$ where ${\cal F}(r_\ins{H})=0$. This point belongs to either outer or inner branch of the apparent horizon. The outgoing null ray, which crosses the first shell at $r_-=r_\ins{H}$, propagates along the horizon, and crosses the second shell at the same radius $r_+=r_\ins{H}$. Consider a narrow beam of outgoing null rays, propagating in the vicinity of this horizon. Denote $y=r-r_\ins{H}$, then in the vicinity of the horizon one has
\be\label{Fy}
{\cal F}={\cal F}'_\ins{H} y+{1\over 2} {\cal F}''_\ins{H} y^2+{1\over 6} {\cal F}'''_\ins{H} y^3+\ldots\, .
\ee
We also denote by $y_{\pm}$ the values of $y$ for the intersection of the null ray with first and second null shell, respectively. Using expression (\ref{bb.2}) one finds
\be
{\cal E}={1\over 192\pi}{{\cal F}'''_\ins{H}\over {\cal F}'_\ins{H}}{y_+^2-y_-^2\over y_+^2}\, .
\ee
In order to establish a relation between $y_-$ and $y_+$, it is sufficient to use the linearized version of the equation (\ref{eq2})
\be
{dy\over dv}={1\over 2} {\cal F}'_\ins{H} y\, .
\ee
Its solution with the initial data $y(0)=y_-$ is
\be\label{yv}
y(v)=y_- \exp(\kappa_\ins{H} v)\, ,
\ee
where $\kappa_\ins{H}={1\over 2} {\cal F}'_\ins{H}$ is the surface gravity at the horizon. Thus one has
\be
y_+=y_- \exp(\kappa_\ins{H} q)\, ,
\ee
and
\be\n{EHH}
{\cal E}_\ins{H}={{\cal F}'''_\ins{H}\over 384\pi  \kappa_\ins{H}}[1-\exp(-2\kappa_\ins{H} q)]\, .
\ee

\section{Standard sandwich model}

\subsection{Metric}

To specify a sandwich model, one needs to specify a static metric between the shells. We start with a simple example. Namely, we put $A=1$ in \eq{a.1}. We assume that $F=1$ outside the interval $(0,\Delta V)$, while inside it has the form
\be\n{a.3}
F=1-{2M R^2\over R^3 +2M \ell^2}\, .
\ee
The corresponding metric was considered by Hayward \cite{Hayward:2005gi} (see also \cite{Frolov:2014jva}). We call this spacetime a {\it standard sandwich model}.

\bigskip

\begin{figure}[tbp]
\centering
\includegraphics[width=6cm]{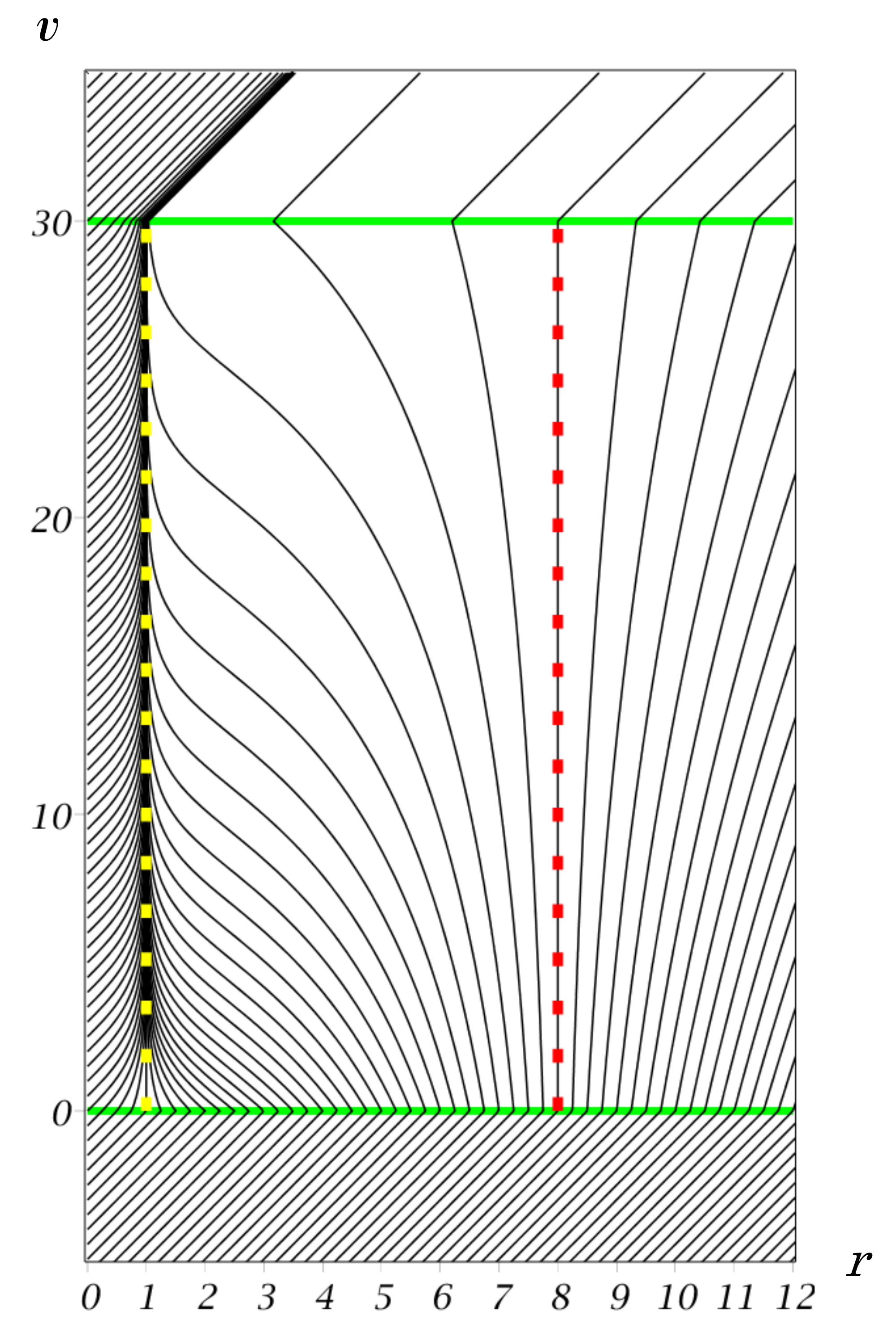}
  \caption{
  This plot shows the outgoing radial null rays $u=\mbox{const}$ propagating in the standard sandwich black hole with $p=8$ and $q=30$.
\label{Orbits_p8q30}}
\end{figure}

The metric function \eq{a.3} contains 2 parameters, the mass $M$ and the "fundamental length" $\ell$. In what follows we assume that the mass $M$ is large enough, so that the function $F$ has two positive zeros $R_{1,2}$. In such a case it is convenient to use $R_{1,2}$ as two new parameters, instead of $M$ and $\ell$. Moreover, we use the radius of the inner horizon $R_{2}$ as a scale parameter $\sigma$. In addition to \eq{sig} we use the following notation
\be
p=R_1/R_2\, .
\ee
It is possible to check that $f(r)$, corresponding to the metric function (\ref{a.3}) has the form
\be\n{a.5a}
f(r)={(r-r_{1})(r-r_{2})(r-r_0)\over r^3 -r_{1} r_{2} r_0},
\ee
where
\be\n{a.5b}
r_{1}=p\hh r_{2}=1\hh r_0=-{p\over p+1}\, .
\ee
The parameter $p$ is a position of the outer horizon in the dimensionless units, while  the inner horizon is located at $r=1$. The quantity $r_0$ is negative. We call it an {\it imaginary horizon}. The metric \eq{a.0},\eq{a.5a} is uniquely specified by two quantities, the mass parameter $p$ and the duration of the black hole existence, $q=\Delta V/R_2$. Because of its rather simple form, a part of the results can be obtained in an analytical form.
The dimensionless surface gravities, calculated for each of these horizons, are
\be\begin{split}\n{a.7}
&\kappa_{1}={(p-1)(p+2)\over 2 p(p^2+p+1)},\\
&\kappa_{2}=-{(p-1)(2p+1)\over 2(p^2+p+1) },\\
&\kappa_0={ (p+1)(p+2)(2p+1)\over 2 p(p^2+p+1)} .
\end{split}\ee

The motion of the incoming radial null rays in this geometry is rather simple. They are described by the equation $v=$const. Such rays pass through the center $r=0$ and become outgoing. Out-going null rays in the standard sandwich black-hole geometry are shown at Fig.\,\ref{Orbits_p8q30}. As we described earlier, there exist three different kinds of these rays. Rays $I$ have their origin as incoming rays with $v<0$. In their propagation they cross both of the shells. Rays that cross the first null shell with $r<1$ propagate near the origin, being accumulated near the inner horizon from its inner side. Rays $I$ that cross the first null shell between $r=1$ and $r=p$ are also attracted to the inner horizon and are accumulated near it from its outer side. Rays $I$ that cross the first shell at $r>p$ are propagating away from the outer horizon. All incoming rays of type $II$, with the origin at $0<v<q$, after passing the inner horizon cross the center $r=0$ and returns to inner horizon from inside. They   are accumulated in the narrow domain in its vicinity. In other words, the inner horizon plays the role of attractor for the outgoing rays, while the outer horizon repulses the rays. After the collapse of the second shell with a negative mass, all the outgoing null rays freely propagate to the future null infinity ${\cal I}^+$.

\bigskip

\begin{figure}[tbp]
\centering
\includegraphics[width=6.5cm]{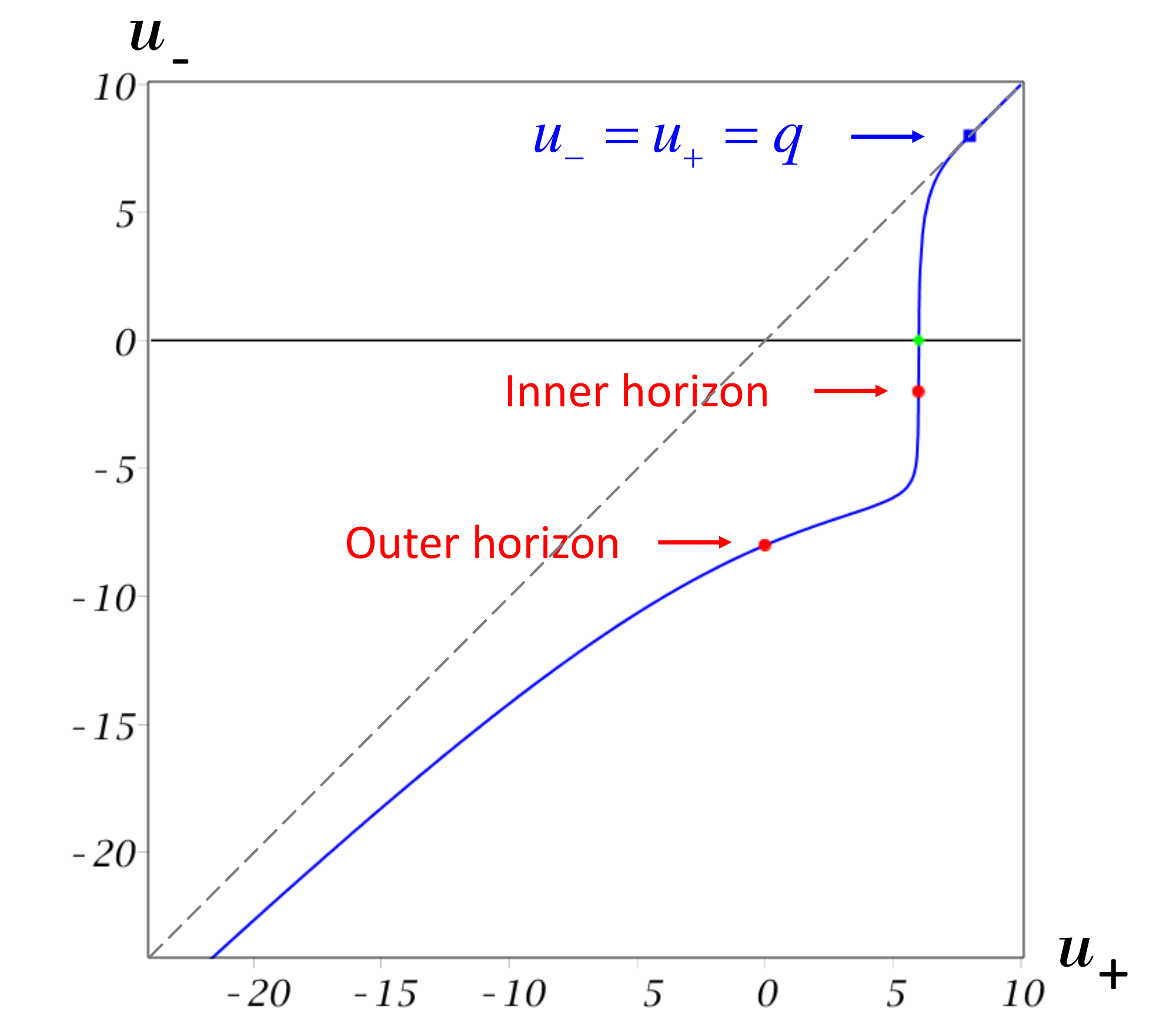}
  \caption{
  This plot shows the function $u_{-}(u_{+})$ for the standard sandwich black hole with $p=4$, $q=8$. The dash line depicts an asymptotic $u_-\to u_+$, when $u_+\to -\infty$ or $u_+>q$.
\label{Uu_p4q8}}
\end{figure}

\bigskip

\begin{figure}[tbp]
\centering
\includegraphics[width=6cm]{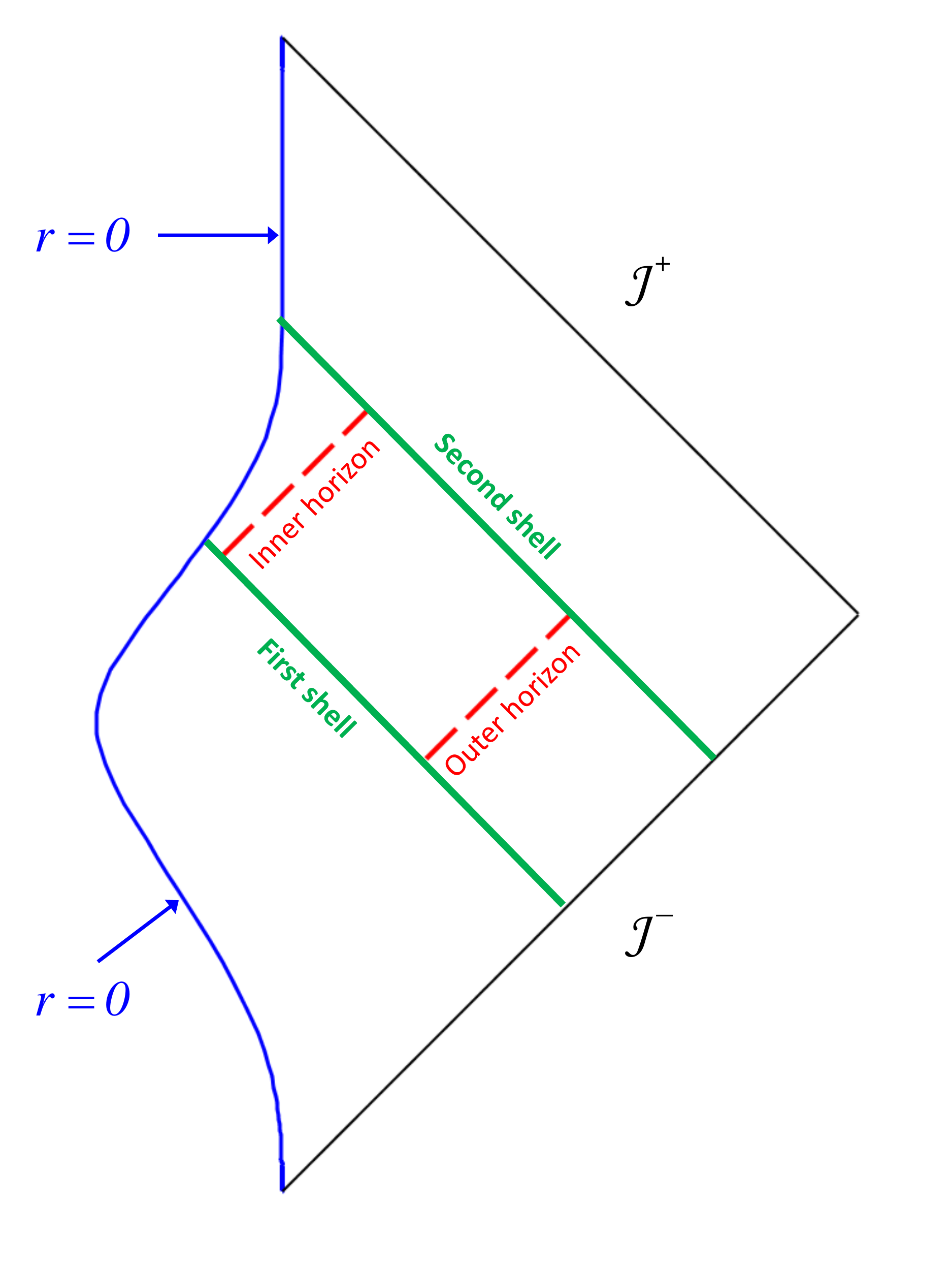}
  \caption
  This  is a Carter-Penrose diagram for the standard sandwich black hole with the parameters $p=3$, $q=3$.
\label{Penrose}
\end{figure}

To establish a relation between the parameters $u_-$ and $u_+$, which is required for the calculation of the quantum energy flux, one needs to calculate the function $Q$, defined by (\ref{calF}), which in the case under considerations takes the form
\be
Q=\int_0^r{dr\over f}\, .
\ee
One has the following expression for $f^{-1}$ in the metric (\ref{a.5a})
\be\n{a.6}
f^{-1}=1+{1 \over2\kappa_{1}( r-r_{1})}+ {1\over 2\kappa_{2}(r-r_{2})}+ {1\over 2\kappa_0(r-r_0)}\, .
\ee
Simple calculation gives
\be\begin{split}\n{a.11}
&Q(r)=r+ {1\over 2\kappa_{2}}\ln |r-1|\\
&+{1\over 2\kappa_{1}}\ln\left({|r-p|\over p}\right)+ {1\over 2\kappa_0}\ln\left({ |r-r_0|\over |r_0|}\right)\, .
\end{split}\ee

\bigskip

\begin{figure}[tbp]
\centering
\includegraphics[width=8.5cm]{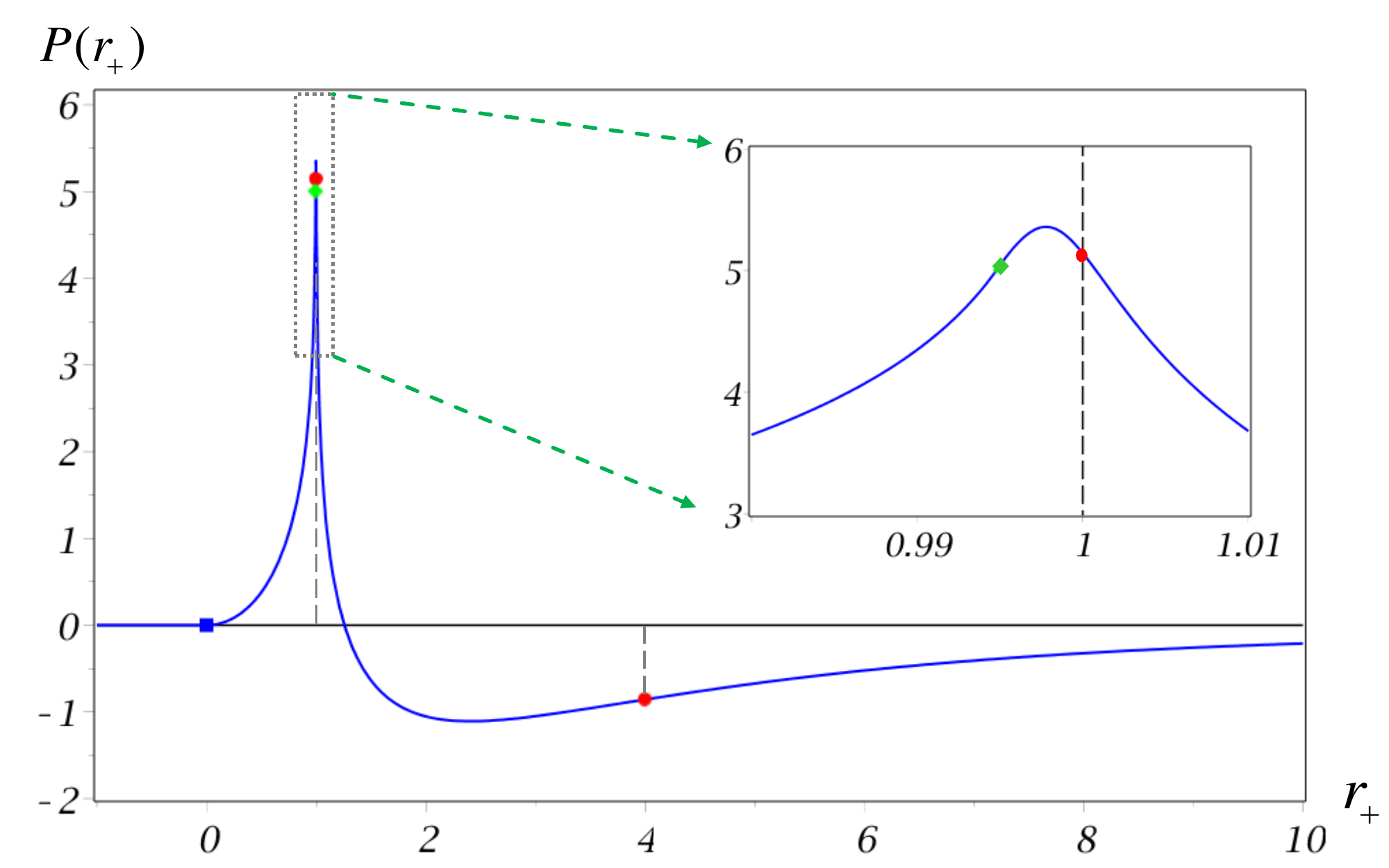}
  \caption{
  This plot illustrates the function $P(r_{+})=\ln \beta$ for the standard sandwich black hole with $p=4$, $q=8$. Two dashed vertical lines represent the position of the inner $r=1$  and outer $r=p$  horizons. The value of  $P$ on the horizons are marked by (red) dots. The square (blue) marks $P$ at the point $r_+=0$. The diamond (green) dot marks $P$ at the point corresponding to $r_{-}=0$.
\label{P_p4q8}}
\end{figure}

Relations (\ref{QQ}) and (\ref{uQQ}) allow one to find relations between $u_+$ and corresponding $u_-$
for type $I$ and $II$ rays, respectively. For $u_->q$, that is for rays of type $III$, one has $u_-=u_+$. Fig.\,\ref{Uu_p4q8} shows this function $u_-(u_+)$for a standard sandwich black hole with special values $p=4$ and  $q=8$. For large negative  $u_+$ the function $u_-(u_+)$ asymptotically tends to the straight line $u_+=u_-$. This region corresponds to the "early" null rays with $v\ll 0$. After passing the center they cross both shells at large radii. In other words, they always propagate in the domain with small (but not vanishing) gravitational potential. This potential is smaller when the value of $-v$ is larger.

The conformal structure of the spacetime of the standard sandwich black hole is shown in the Carter-Penrose diagram Fig.\,\ref{Penrose}. To compactify the null coordinates $(u,v)$ we used new coordinates $\bar{u}=\arctan(u/\gamma)$ and $\bar{v}=\arctan(v/\gamma)$ and corresponding Cartesian spacetime coordinates $\zeta=\bar{v}-\bar{u}$ and $\eta=\bar{v}+\bar{u}$. The parameter $\gamma$ is an arbitrary positive constant. We chose $\gamma=3$, for the better presentation of the diagram. In these coordinates
null rays $u=\const$ and $v=\const$ are represented by straight lines with the
slope $\pm 1$. The future null infinity ${\cal I}^+$ and the past null infinity ${\cal I}^-$ are given by segments of the lines $\eta+\zeta=\pi$ and $\eta-\zeta=\pi$. At both null infinities the asymptotic Killing vector is normalized to unity. The solid (green) lines on the Fig. \ref{Penrose} depict the shells of infalling null matter at $v=0$ and $v=q$. The dashed (red) lines between them correspond to the inner and outer horizons of the standard sandwich black hole. The curve corresponding to the center of the black hole $r=0$ can be calculated using the function $u_{-}(u_{+})$ (see section \ref{Static}). The upper part of this curve (above the second shell) is a vertical straight line.

Fig.\,\ref{P_p4q8} shows the logarithm of the gain function $P(r_+)=\ln \beta$. This function has a peak in the vicinity of the inner horizon, though the maximum of the gain function is not exactly on the inner horizon. For the given parameters of the standard sandwich black hole $p=4$, $q=8$ this maximum is located slightly below the horizon. Note that in a generic case the maximum can be either below or above the inner horizon. For the black holes with large lifespan $q\gg |\kappa_2|^{-1}$ the peak is exponentially narrow $\sim \exp(-|\kappa_2|q)$ and looks very sharp, but it is, in fact, a smooth function on the top. Its amplitude is of the order of $|\kappa_2|q$.

\subsection{Hawking radiation}

We demonstrate now that Hawking result on the quantum energy flux from a black hole is correctly reproduced when the mass parameter $p$ and the lifespan $q$ are large. In this case the metric function outside the external horizon $f$ can be approximated as follows
\be\n{pS}
f=1-{p\over r}\, ,
\ee
and one has
\be\n{kS}
\kappa={p\over 2r^2}\hh R={2p\over r^3}\hh B={p(4r-3p)\over r^4}\, .
\ee
The null rays with $r_+>p$ cross both shells.
Let us denote $Y=y +\ln(y -1)$.
Then, in the adopted approximation, the relation (\ref{QQ}) takes the form
\be\n{YY}
Y(y_+)={q\over 2p}+Y(y_-)\hh y_{\pm}={r_{\pm}\over p}\, .
\ee
The function $Y$ vanishes at $y=y_*$, where
\be
y_*=W(e^{-1})\approx 1.27846\, .
\ee
where $W(z)$ is a Lambert-W function.
If $y_+<y_*$ then $Y(y_+)<0$ and \eq{YY} implies that $Y(y_-)<0$. That is both $r_+$ and $r_-$ are close to the outer horizon, and the energy flux connected with this beam of null rays can be estimated by using \eq{EHH}. Let us consider the case when $y_+\gg y_*$. In this domain $Y(y_+)$ can be approximated as follows $Y(y_+)\sim y_+$ and one has
\be
B(y_+)p^2\ll 1\hh f(y_+)\approx 1\, ,
\ee
so that the relation (\ref{bb.2}) can be approximated as follows
\be\n{c.2}
{\cal E}={B(r_-)\over 192\pi}\, .
\ee
If $Y(y_+)<q/(2p)$ than \eq{YY} shows that $y_-<y_*$ and $B(r_-)\approx p^{-2}$. When $Y(y_+)\gg q/(2p)$ the quantity $B(r_)$ is small. To summarize, in the interval $r_+\in (y_* p, q/2)$ the energy flux is approximately constant
\be\n{EHHH}
{\cal E}={\cal E}_\ins{Hawk}={\kappa^2\over 48\pi}={1\over 192\pi p^2}\, ,
\ee
and for $r_+>q/2$ it quickly decreases and vanishes. The expression (\ref{EHHH}) correctly reproduces the result for the Hawking flux.

\bigskip

\begin{figure}[tbp]
\centering
\includegraphics[width=8cm]{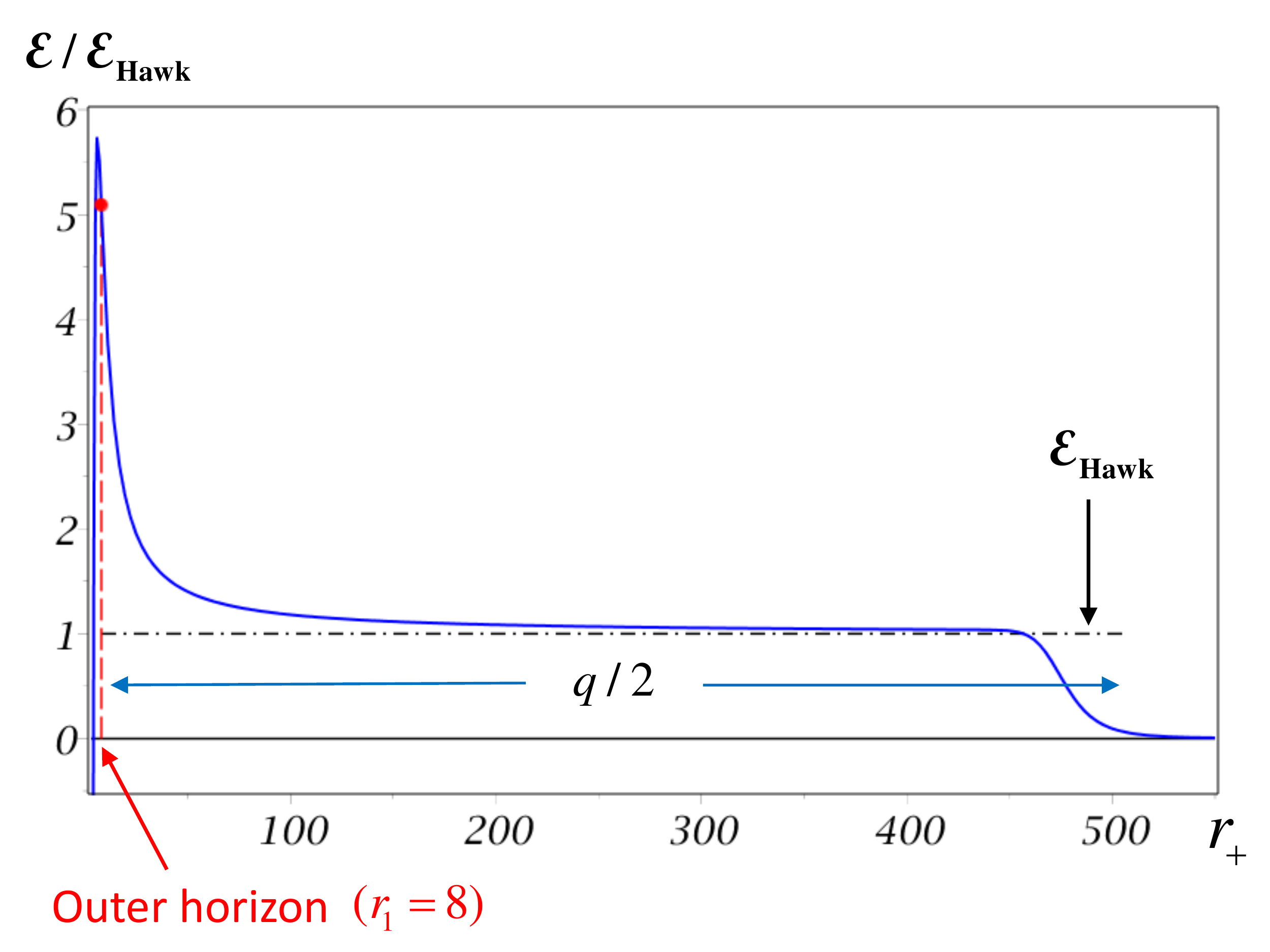}
  \caption{
  This plot depicts the flux of particles emitted by the standard sandwich black hole with $p=8$, $q=1000$ at the moment $v=q$ in the range of radii from the outer horizon $r_1=8$ till infinity. The dash line denotes the asymptotic of the Hawking radiation in the limit of $q\to \infty$. At radii greater than $\approx q/2$ the flux vanishes. It corresponds to the coordinate $u+{-}\approx 0$. Thus the tail of the Hawking radiation lasts $\approx q$.
\label{Flux_p8q1000}}
\end{figure}

Fig.\,\ref{Flux_p8q1000} shows the result of the numerical calculation of the energy flux from the standard sandwich black-hole in its external domain. A peak near the outer horizon describes emission of photons, created near the outer horizon, which are ''released'' when the second shell crosses it. The properties of this part of radiation are model-dependent. However, soon after this, the flux of the photons is stabilized and reaches the values ${\cal E}_\ins{Hawk}$, given by (\ref{EHH}).
When $q\gg p$ the duration of the phase of the Hawking radiation is $\Delta u\approx q$ and the total energy emitted during this phase is $q/(192 \pi p^2)$. It is instructive to present the obtained results in the dimensional form. We remind, that we put $\sigma=r_-\approx \ell$. The duration of time of the Hawking radiation phase is $\Delta U=\Delta V$ and the rate of the energy emission, $\dot{E}$ and total emitted energy, $\Delta E$, are
\ba
&&\dot{E}={\ell_{Pl}^2\over 192 \pi r_g^2}={\pi\over 12} (kT)^2\hh kT={1\over 4\pi r_g}\, ,\n{ETT}\\
&&\Delta E={\pi\over 12} (kT)^2 \Delta V\, .\n{DEE}
\ea
Relation (\ref{ETT}) correctly reproduces the result of \cite{Christensen:1977jc} for the 2D energy flux, calculated from the trace anomaly.

Relation (\ref{DEE}) implies that the total energy of the Hawking radiation emitted by a standard sandwich black hole is proportional to the duration of its existence, $\Delta V$. This is a natural result, which clearly demonstrates that the emitted in this process quanta are created with a constant rate near the horizon \footnote{Sometimes one can see in the literature a statement that all the particles of the Hawking radiation are created during the formation of the black hole. The relation (\ref{DEE}) shows that such a statement is totally wrong. A decision when to switch-off the black hole can be made any time after its formation, so that $\Delta V$ is an arbitrary parameter.}.

\subsection{Quantum radiation from the inner horizon}\label{QuantumRadiation}

As a result of a focusing property of the inner horizon, the beam of the outgoing null rays crossing the second null shell for small $r_+\ll p$ is sharply concentrated near $r_+=1$. For this reason for the calculation of the quantum radiation from the inner horizon one can use the following approximation
\be\begin{split}\n{c.9}
Q(r_{-})&={1\over 2\kappa_{1}}\ln\left({|r_{-}-p|\over p}\right)+r_{-}\\
&+ {1\over 2\kappa_{2}}\ln |r_{-}-1|+ {1\over 2\kappa_0}\ln\left({ |r_{-}-r_0|\over |r_0|}\right),\\
Q(r_{+})&={1\over 2\kappa_{2}}\ln\left(|r_{+}-1|\right)+1\\
&+ {1\over 2\kappa_{1}}\ln  \left({|p-1|\over p}\right) + {1\over 2\kappa_0}\ln\left({ |1-r_0|\over |r_0|}\right).
\end{split}\ee
Consider type $I$ rays. Solving equation (\ref{QQ}) one finds $r_+$ as a function of $r_-$. Using the notations
\be\begin{split}\n{c.10}
&y_+=r_+ -1\hh y=r_{-} -1,\\
& {\cal B}=e^{-\kappa_{2} q}\hh Y={\cal B}y_+\, ,
\end{split}\ee
one obtains
\be\begin{split}\nonumber
&Y= Z(y) ,\\
&Z(y)=y e^{2\kappa_{2} y}
\left({p -1-y\over p-1}\right)^{\kappa_{2}/\kappa_{1}}\left({1+y -r_0\over 1-r_0}\right)^{\kappa_{2}/\kappa_0} ,\\
& {dr_-\over dr_+}={{\cal B}\over Z'}\hh P=-\ln |Z'|+\ln {\cal B} ,\\
& {dP\over dr_+}=-{\cal B}{Z''\over Z'^2}\hh
{d^2P\over dr_+^2}=-{\cal B}^2\left( {Z'''\over Z'^3}-2{Z''^2\over Z'^4}\right) ,\\
& {\cal E}={{\cal B}^2\over 192 \pi} \left(2{Z'''\over Z'^3}-3{Z''^2\over Z'^4}\right) .
\end{split}\ee
Let us notice, that the parameter $q$ enters the expression for ${\cal E}$ only in the combination ${\cal B}=e^{-\kappa_{2} q}$, which is a common scaling factor of transformation between $r_-$ and $r_+$ coordinates. Namely, after re-scaling $r_+-1={\cal B}^{-1}Y$ the expression for ${\cal B}^{-2}{\cal E}$ as a function of $Y$ has a universal behavior independent of $q$. Since the surface gravity of the inner horizon, $\kappa_{2}$, is negative the scaling factor ${\cal B}$ for large $q$ is exponentially large. This fact reflects the following general property of the radiation from the inner horizon. Quanta, propagating near it, experience huge blueshift and ${\cal B}$ is the corresponding blueshift factor. Using this fact one can estimate the total energy of the radiation, emitted from the inner horizon as follows
\be
\Delta E \sim {\cal E} \Delta u_+\sim {\cal B} \Delta E_0\, ,
\ee
where $\Delta E_0$ is a scale-invariant value of the energy, which is of the order of 1.

\subsection{Numerical results}

\bigskip

\begin{figure}[tbp]
\centering
\includegraphics[width=7cm]{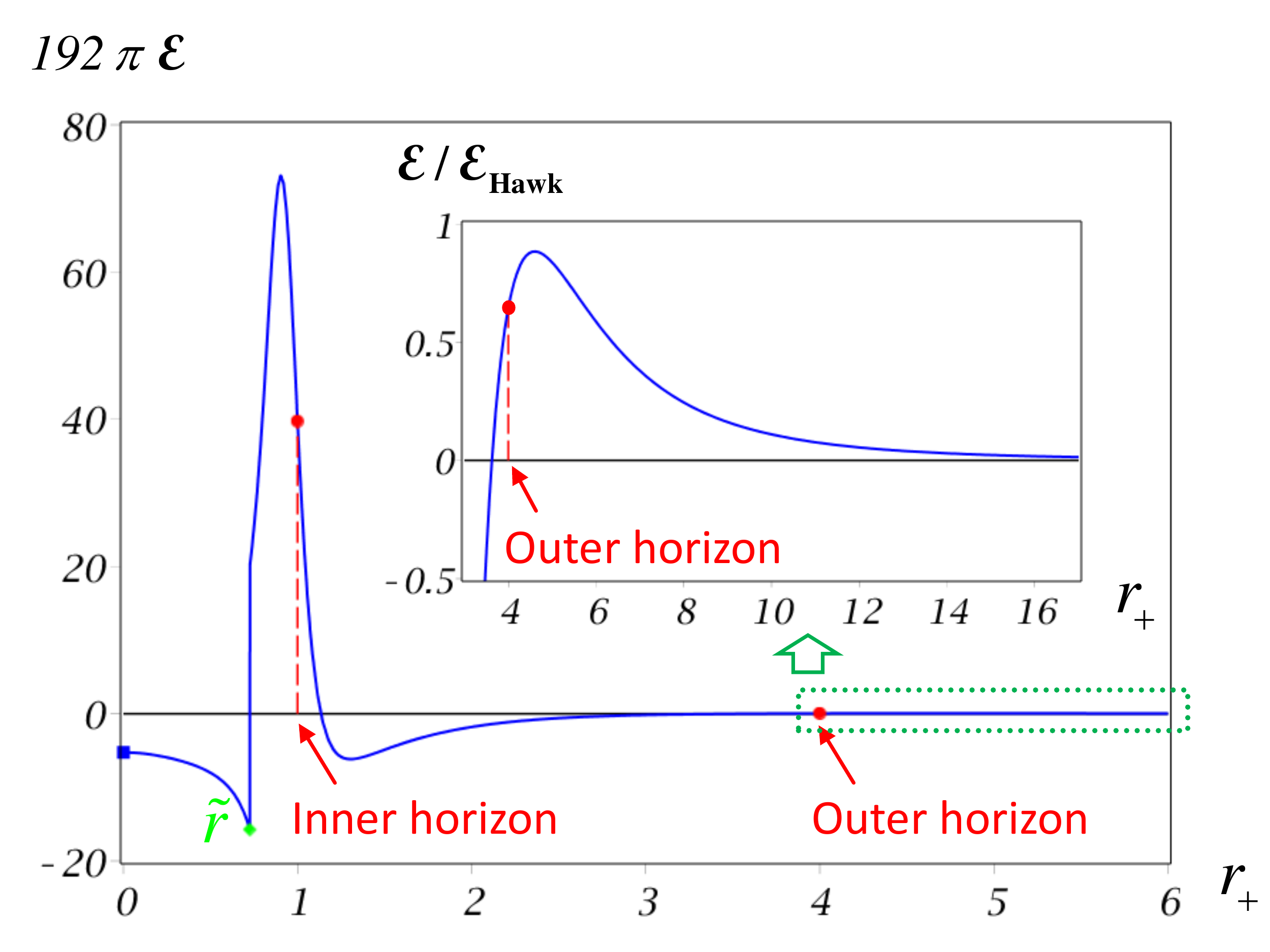}
  \caption{
  This plot depicts the flux of particles emitted by the standard sandwich black hole with $p=4$, $q=2$ at the moment $v=q$ in the range of radii from $r=0$ till $r=16$.
\label{Flux_p4q2}}
\end{figure}

\bigskip

\begin{figure}[tbp]
\centering
\includegraphics[width=7cm]{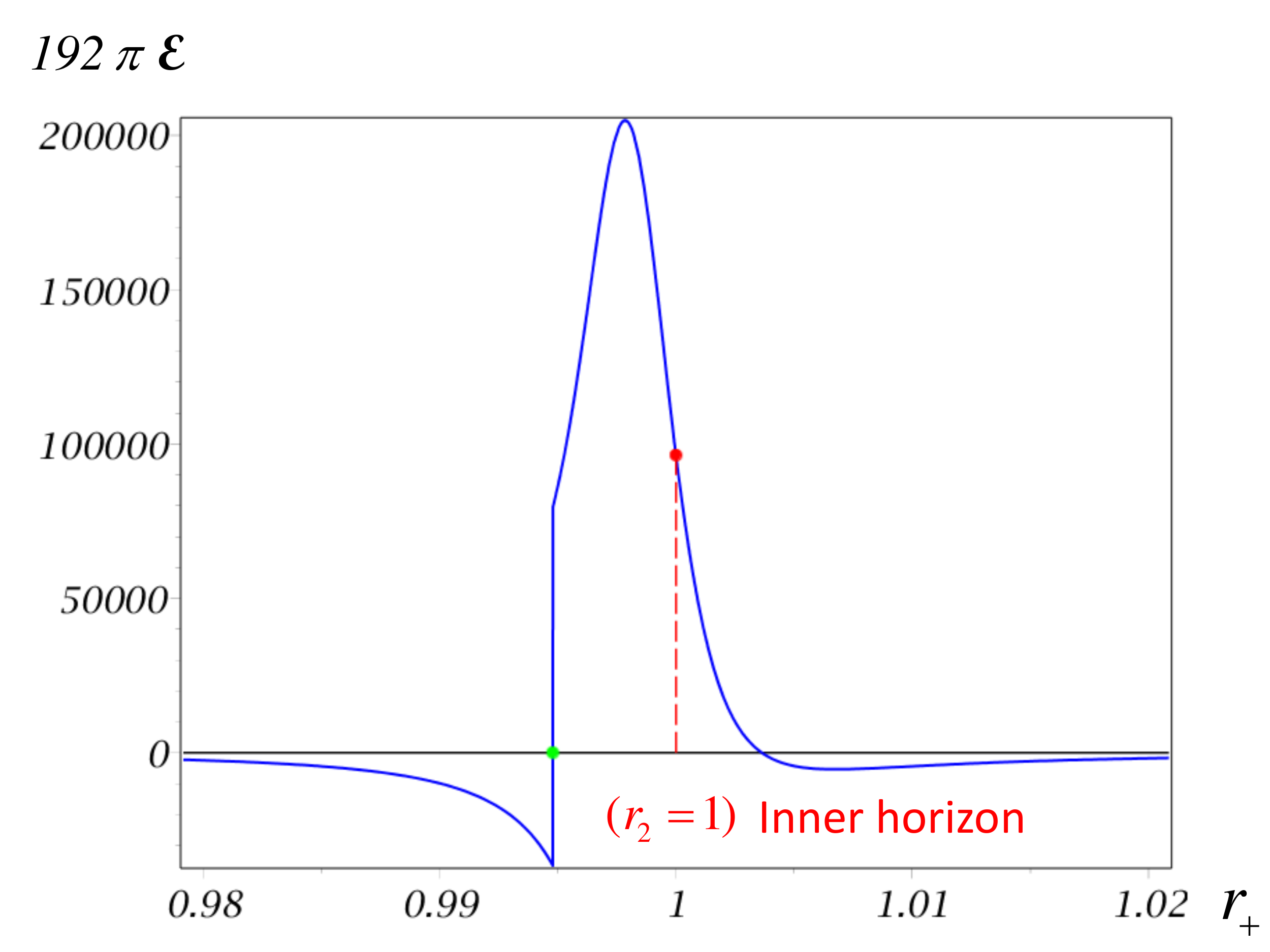}
  \caption{
  This plot depicts the flux of particles emitted by the standard sandwich black hole with $p=4$, $q=8$ at the moment $v=q$ in the range of radii in the close vicinity of the inner horizon $r_2=1$. The maximum amplitude of the radiation grows exponentially with  $q$ as $\exp({-2 \kappa_2 q})$, while the width of the peak shrinks as $\exp({\kappa_2 q})\ll1$ (Note that $\kappa_2$ is negative).
\label{Flux_p4q8a}}
\end{figure}

\bigskip

\begin{figure}[tbp]
\centering
\includegraphics[width=7cm]{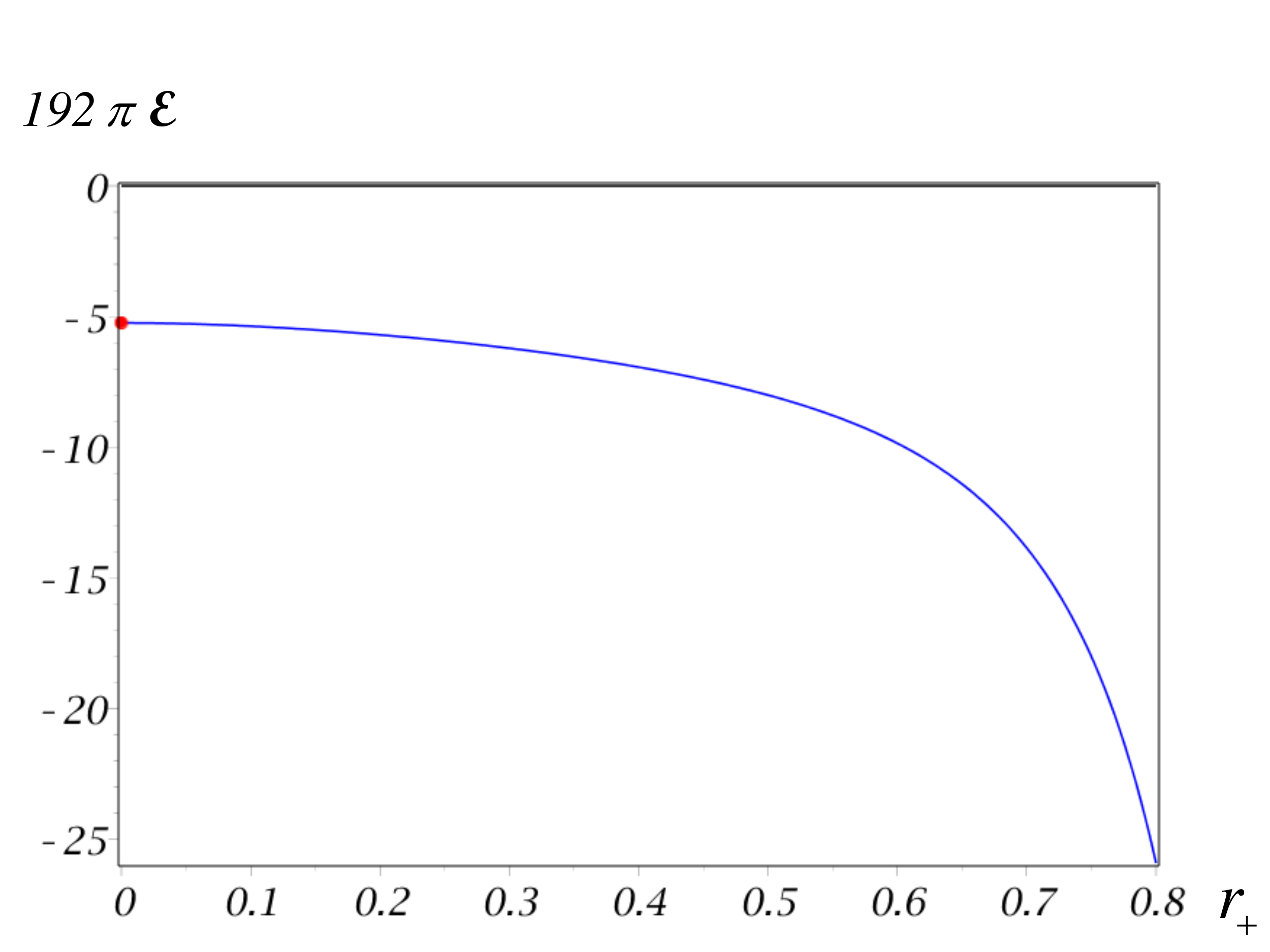}
  \caption{
  This plot depicts the flux of particles emitted by the standard sandwich black hole with $p=4$, $q=8$ at the moment $v=q$ in the range of radii inside the inner horizon from the center $r=0$ to $r=0.8 r_2$. For any given parameter $p$ the shape of the curve does not depend on $q$ till the radius $\tilde{r}=r_{+}(0)$. The position of this point depends on q and at large $q$ we have $r_2-\tilde{r}\approx \exp({\kappa_2 q})\ll 1$.
\label{Flux_p4q8b}}
\end{figure}

We already mentioned that the numerical calculations for large $p$ and $q$ are in agreement with the Hawking result for standard sandwich black holes, provided the duration parameter $q$ is large enough. Let us discuss now quantum radiation of the standard sandwich black hole in other domains. Fig.\,\ref{Flux_p4q2} shows the energy flux from a standard sandwich black hole with parameters $p=4, q=2$. For such small value of the duration parameter $q$ there is no enough time to develop constant Hawking radiation. At the moment of time when the outer horizon crosses the second shell there exists a flash of the radiation, which changes its sign from positive (out the horizon) to negative (in the domain inside the outer horizon). The radiation emitted between the outer- and inner-horizons remains relatively small. The very intensive outburst of the energy occurs near the inner horizon. We choose this small value of $q$ just to be able to show on the plot the radiation from all the domains. As we already mentioned in the previous section, for higher value of the duration parameter $q$ the amplitude of the burst radiation grows as $\sim \exp(-2\kappa_2 q)$, while the width of the peak decreases as $\sim \exp(\kappa_2 q)$. Fig.\,\ref{Flux_p4q8a} shows the energy burst from the inner horizon for the standard sandwich black hole with parameters $p=4, q=8$. One can also see in this figure that for the radii smaller that $1$ the energy flux become negative. This region in more details is shown in Fig.\,\ref{Flux_p4q8b}. An analysis shows that this region is connected with type $II$ rays.

Let us summarize. Till now we studied what was called a standard sandwich black-hole model. In such a model the metric between two null shell was chosen to coincide with the Hayward metric \cite{Hayward:2005gi}. A characteristic property of this geometry is that $\alpha=1$, so that a falling photon, when it reaches the center, $r=0$, has the same energy, as at the infinity. In other words, there is no red- or blueshift for such photons. One of the consequences of this assumption is that the surface gravity at the inner horizon is high, and for large mass parameter $p$ is $\kappa_2\approx -1$. As we demonstrated the quantum radiation from the inner horizon of such a black hole is  high. For large duration parameter, $q$, the energy emitted from it is proportional to $\exp(2 q)$ and easily exceeds the mass of the black hole $\sim p/2$. This property shows that such  standard models are internally inconsistent.

Certainly, the standard sandwich model is quite different from a ``realistic'' black hole, where the mass decrease is not abrupt, but is a smooth and continuous function of $v$. However one can expect that the main conclusion, concerning quantum inconsistency of such a model, still remains valid. A reason for this is the following. The effect of the huge energy outburst from the inner horizon is deeply connected with the formula (\ref{betagen}) for the energy gain parameter $\beta$. For the Hayward metric $\alpha_0=1$ and $\pa_r(f)|_{f=0}$ at the inner horizon is negative. For a static metric this quantity is negative and of the order of $-1$ (in the adopted units). This property remains valid during almost all of the time for an evaporating black hole, provided its initial mass is much larger than the Planckian one.

\subsection{Possible role of quantum fluctuations}

In the derivation of this result we assumed that the inner horizon is an infinitely sharp surface. One might assume that quantum fluctuations, that smear the horizon, can dramatically modify the expression for the flux, emitted from the inner horizon. Let us discuss this option and demonstrate such a mechanism of the energy outburst suppression apparently does not work. In order to discuss a possible role of quantum fluctuations one can use formalism, developed in \cite{Barrabes:1998iw,Barrabes:2000fr}. Namely, let us assume that the mass parameter $M$ in the metric function (\ref{a.3}) fluctuates and it is of the form
\be
M=M_0+\mu(v)\hh \mu(v)=\mu_0 \cos(\omega v +\phi)\, .
\ee
As a result, the map between ${\cal I}^{-}$ and ${\cal I}^+$, which is determined by the radial null rays, would depend on $\mu(u)$. To describe a fluctuating horizon one should consider the phase $\phi$ and the amplitude $\mu_0$ as stochastic variables. As a result of averaging, the position of the horizon becomes uncertain, and the horizon  is effectively smeared. The characteristic width of this broadening is determined by the average value of $\mu_0$.

Let us consider a classical massless particle, propagating in the vicinity of the inner horizon, and find how its gain function is modified under the action of the fluctuations. For estimation of the gain function we use the expression (\ref{betagen}). In our case $\alpha=\alpha_0=1$. Let us expand function $f$ in the vicinity of the horizon
\be
f\approx \kappa_2 (r-1)+\left. \pa_M f\right|_{r=1} \mu(v)\, .
\ee
Thus
\ba
&&\beta \approx \exp(-\kappa_2 q)\exp  b\, ,\\
&& b= -{1\over 2\omega} \left. \pa_r\pa_M f\right|_{r=1}\mu_0  [\sin(\omega q+\phi)-\sin(\phi)]\, .
\ea
For a small amplitude $\mu_0$ one gets $\langle \exp b\rangle \approx 1+O(\langle \mu_0\rangle^2)$. Thus small fluctuations of the horizon only slightly modify the gain function and its leading term has the same form $\sim \exp(-\kappa_2 q)$ as in the absence of fluctuations. This implies that the same  conclusion should be valid for the value of the outburst radiation from the inner horizon in this model.

\section{Quantum radiation from a modified sandwich black hole}

\subsection{Modified model}

We consider now a modified version of the sandwich black-hole, where redshift factor $\alpha$ is present. Namely, we consider a double shell model and choose the modification \cite{Frolov:2016pav} of the Hayward metric between the null shells in the form
\be\begin{split}\label{dsalpha}
&ds^2=-\alpha^2 f dv^2+2\alpha dv dr +r^2 d\omega^2\, ,\\
&f={(r-p)(r-1)(r+{p\over p+1})\over r^3+{p^2\over p+1}}\, ,\\
&\alpha={r^n+1\over r^n+1 +p^k}\, .
\end{split}\ee
 Similar static modification of the Hayward metric with a time delay at the center was proposed in \cite{DeLorenzo:2014pta}. It was shown there that
to be physically plausible, regular black hole metrics
should  incorporate a non-trivial
time delay between an observer at infinity and an observer in the regular center. Generalization to the rotating case was also considered in \cite{DeLorenzo:2015taa}.

To stress that this model differs from the standard one by the presence of the redshift function $\alpha$ we call it an {\it $\alpha$-sandwich model}. As earlier, we assume that $p$ is large.
In order to preserve the correct Schwarzschild asymptotic form of the metric one must put $n\ge 2$. To preserve the value of the surface gravity of the outer horizon at the level of the order $p^{-1}$, one must have $n\ge k+1$. The surface gravity of the inner horizon is of the order of $\kappa_2\sim -p^{-k}$. If we assume that $q$ is of the order the Hawking evaporation time, $p^3$, then the ''dangerous'' blueshift factor, which enter the expression for the rate of the energy emission by the inner horizon, $\exp(-2\kappa_2 q)$ becomes $\sim \exp(2 p^{3-k})$. For large black holes $p\gg 1$ this factor does not grow with the black hole mass, if we assume that $k\ge 4$. To be more specific we present calculations for the a special case $(k=4,n=6)$.

\bigskip

\begin{figure}[tbp]
\centering
\includegraphics[width=6cm]{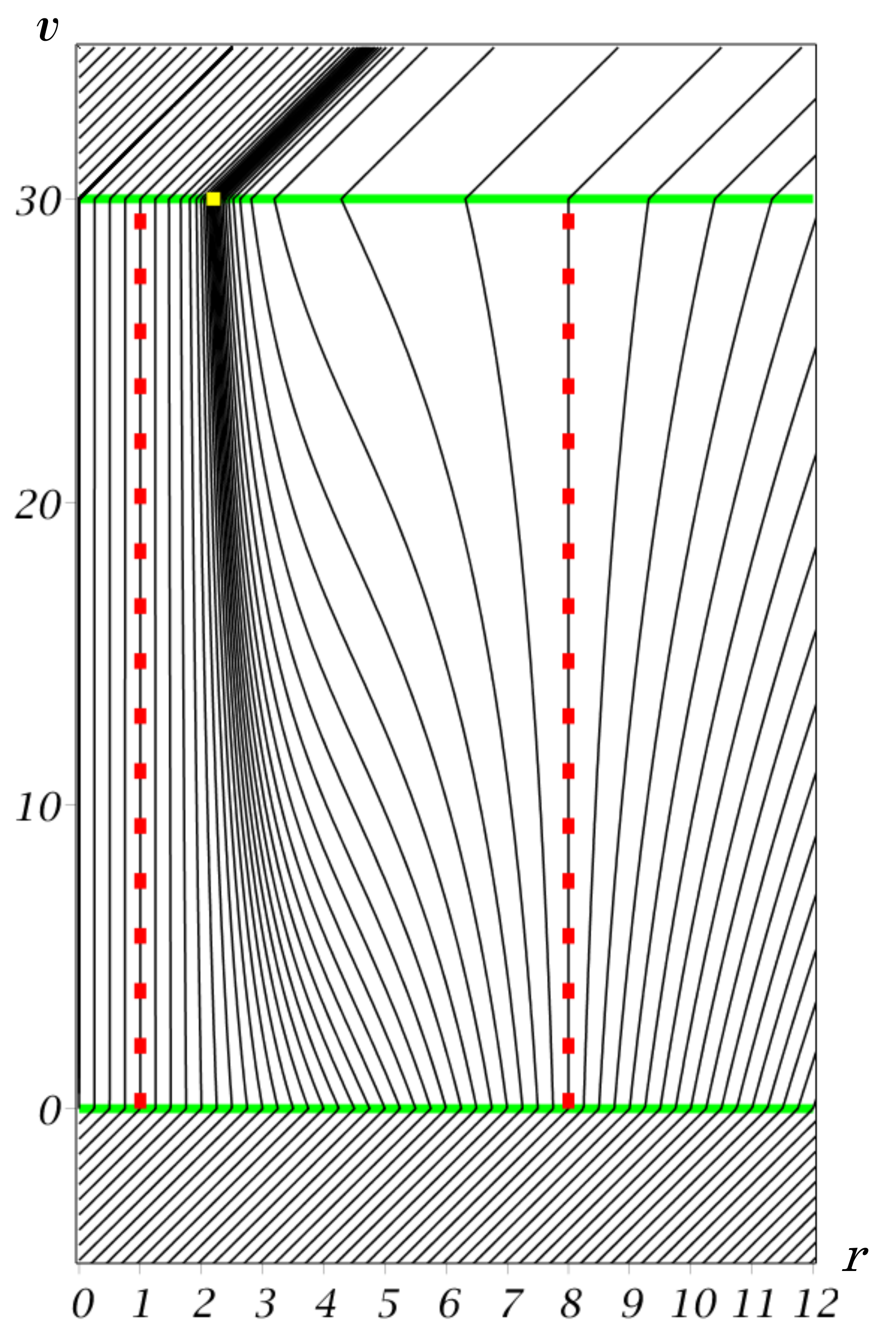}
  \caption{
  This plot shows the null rays $u=\mbox{const}$ propagating in the $\alpha$-sandwich black hole with $p=8$, $q=30$, and  $k=4, n=6$.
\label{Orbits_p8q30k4}}
\end{figure}

Fig.\,\ref{Orbits_p8q30k4} shows the outgoing null rays in the $\alpha$-sandwich black hole geometry. Incoming radial null rays, as earlier, are described by the equation $v=$const. Comparison with Fig.\,\ref{Orbits_p8q30} shows that the attraction of the rays to the inner horizon weakened. This is a reflection of the fact, that  the absolute value of the surface gravity $\kappa_2$ is smaller, than in the standard model.  For the chosen value of $q$ the outgoing rays are accumulated at the radius $r \approx 2.1969$ ( in Fig.\,\ref{Orbits_p8q30k4} it is marked on the second shell by the yellow box). In this aspect quantum radiation of $\alpha$-sandwich black hole differs from that of the standard one, where peak of radiation is located in the close vicinity of the inner horizon.
For larger $q$ the accumulation point shifts closer to the inner horizon. Note that typically the width of this accumulation region is larger than in the standard case.

\subsection{Accumulation effect}

The difference between the  $\alpha$-sandwich black hole and the standard one comes from the effect of the $\alpha$ function on the pace of time between the horizons. The choice of the smooth function $\alpha$ in \eq{dsalpha} guarantees that it has little effect on the properties of the $\alpha$-sandwich black hole near and above the outer horizon, but it considerably slows down time below some radius between $r_2$ and $r_1$. As a consequence the surface gravity of the inner horizon is much smaller than that of the outer horizon.

\bigskip

\begin{figure}[tbp]
\centering
\includegraphics[width=7cm]{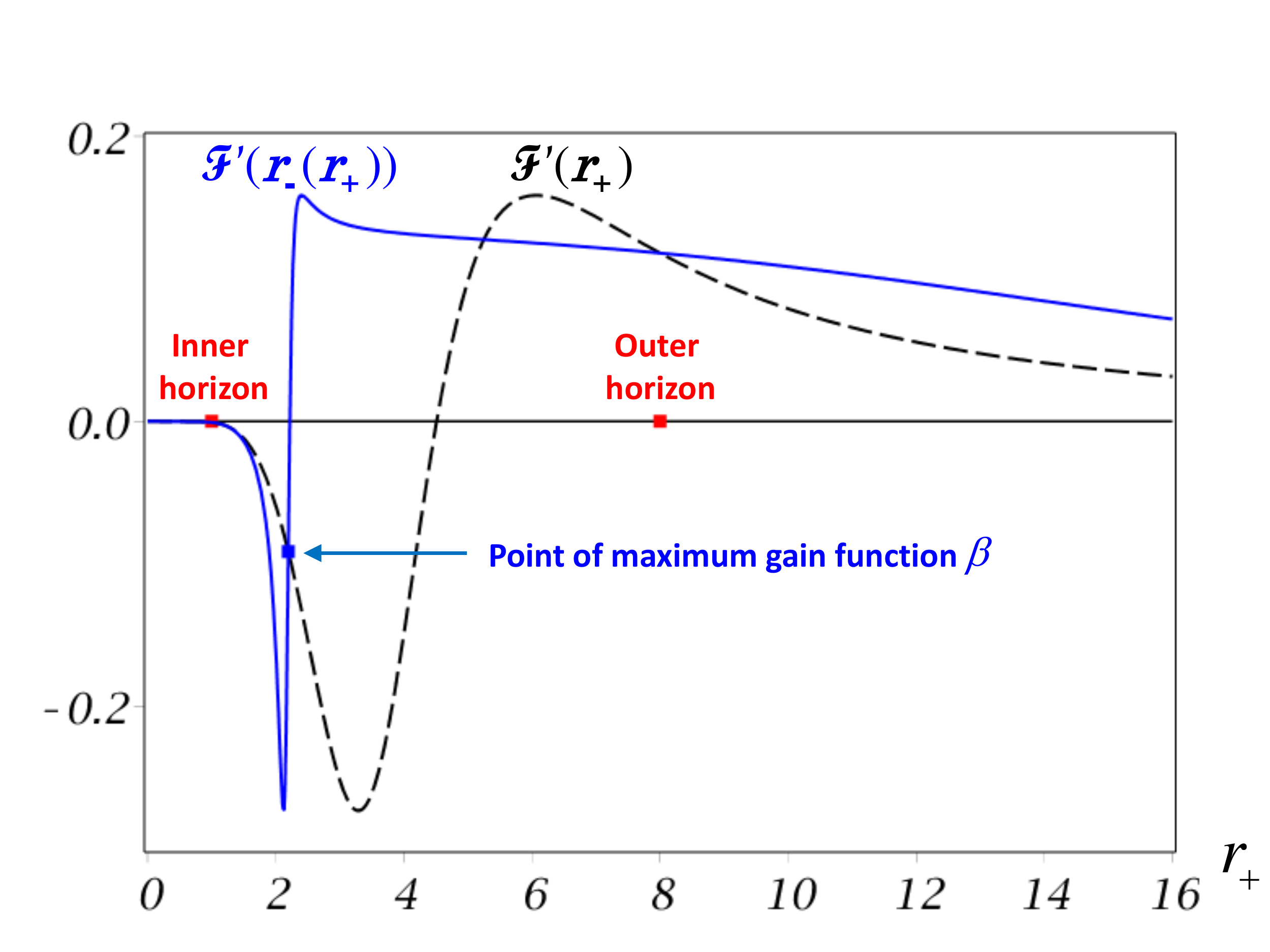}
  \caption{
This plot depicts the functions ${\cal F}'(r_{+})$ (black dashed line) ${\cal F}'(r_{-}(r_{+}))$ (blue solid line) for the $\alpha$-sandwich black hole with $p=8$, $q=30$, and  $k=4, n=6$. The maximum gain function $\beta$ is achieved at the point, where they cross each other at negative values. For the given parameters this point is at $r_{+}=\rho\approx 2.1969$.
\label{rtop_p8q30k4}}
\end{figure}

The effect of the accumulation of the outgoing null rays has a simple explanation. Denote as usual by $r_{\pm}$ the radii of intersection of a null ray with null shells. The rays emitted in the interval $\Delta r_-$ from the first shell cross the second shell in the interval $\Delta r_+$. The gain function $\beta$, which is the ratio $\Delta r_-/\Delta_+$, describes the compression of the beam of the null rays. The higher $\beta$ the larger is the compression. Hence,
the radius $r_+$, where the outgoing rays are accumulated, corresponds to the ray with the maximum gain function $\beta(r_+)$. From the definition of the gain function \eq{beta} it follows that the maximum is achieved when
\be
{d\over dr_{+}}\beta={d\over dr_{+}}\left({ {\cal F}(r_{-})\over {\cal F}(r_{+})}\right)=0.
\ee
It is easy to check that this condition is satisfied when the $r_-$ and $r_+$ are connected by the relation
\be\label{dF}
{\cal F}'(r_{-})={\cal F}'(r_+)\, .
\ee

The behavior of the function ${\cal F}'(r)$ is qualitatively similar for all sandwich black holes in question. It is presented in Fig.\,\ref{rtop_p8q30k4}. From the symmetry considerations it is evident that the function ${\cal F}'(r)$ vanishes at $r=0$. Then it becomes negative. It has  the value $2\kappa_2$, which is also negative, on the inner horizon. It reaches minimum and then it increases. ${\cal F}'(r)$ changes its sign at some radius between the horizons. On the outer horizon it is positive and equals to $2\kappa_1$. Finally it asymptotically vanishes at large radii. Because of such a behavior this function has the following property. Choose a point $r=\rho$ in the vicinity of the inner horizon, where the slope ${\cal F}''(\rho)$ is negative. Then there exists a point $r=\zeta(\rho)<r_2$, where the slope ${\cal F}''(\zeta(\rho))$ is positive, such that the following relation is valid
\be\n{FFp}
{\cal F}'(\rho)={\cal F}'(\zeta(\rho))\, .
\ee
We call $\zeta(\rho)$ a point conjugated to $\rho$. In order that the top of the gain function at $v=q$ be located at $\rho$, the null ray passing throw it must be emitted at $v=0$ at $\zeta(\rho)$. That is the following condition must be satisfied
\be\label{qF}
q=2\int_{\rho}^{\zeta(\rho)} {dr\over|{\cal F}(r)|}.
\ee
The relations (\ref{FFp}) and (\ref{qF}) can be used to determine $\rho=\rho(q)$ for the position of the accumulation point at time $q$.

\bigskip

\begin{figure}[tbp]
\centering
\includegraphics[width=7cm]{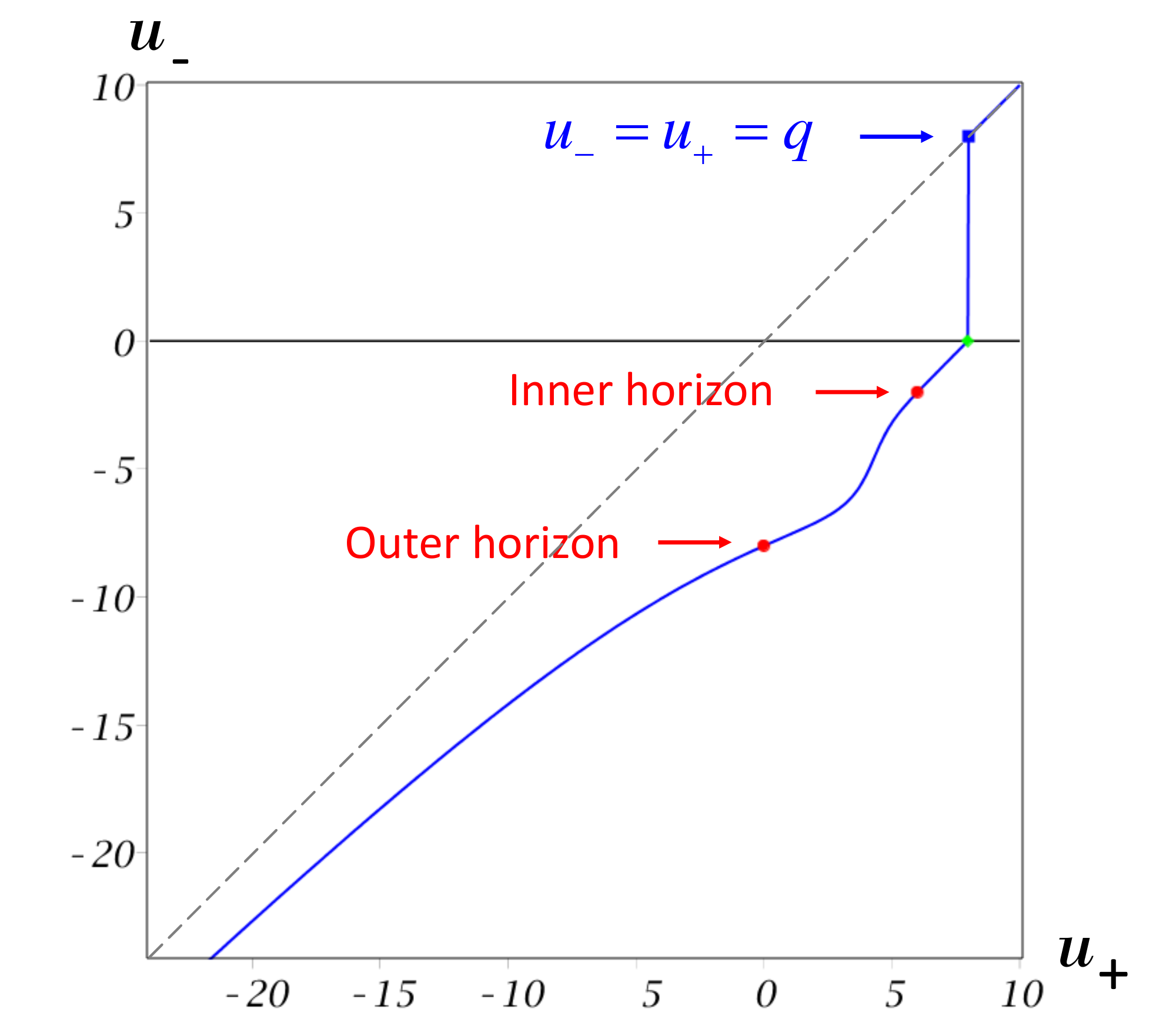}
  \caption{
  This plot shows the function $u_{-}(u_{+})$ for the $\alpha$-sandwich black hole with $p=4$, $q=8$, and $k=4, n=6$. The dash line depicts an asymptotic $u_-\to u_+$, when $u_+\to -\infty$ or $u_+>q$. The curve connecting the green and the blue dots,  which mark points $u_{-}=0$ and $u_{-}=u_{+}=q$ correspondingly, is very close to a vertical line.
\label{Uu_p4q8k4}}
\end{figure}

\bigskip

\begin{figure}[tbp]
\centering
\includegraphics[width=6cm]{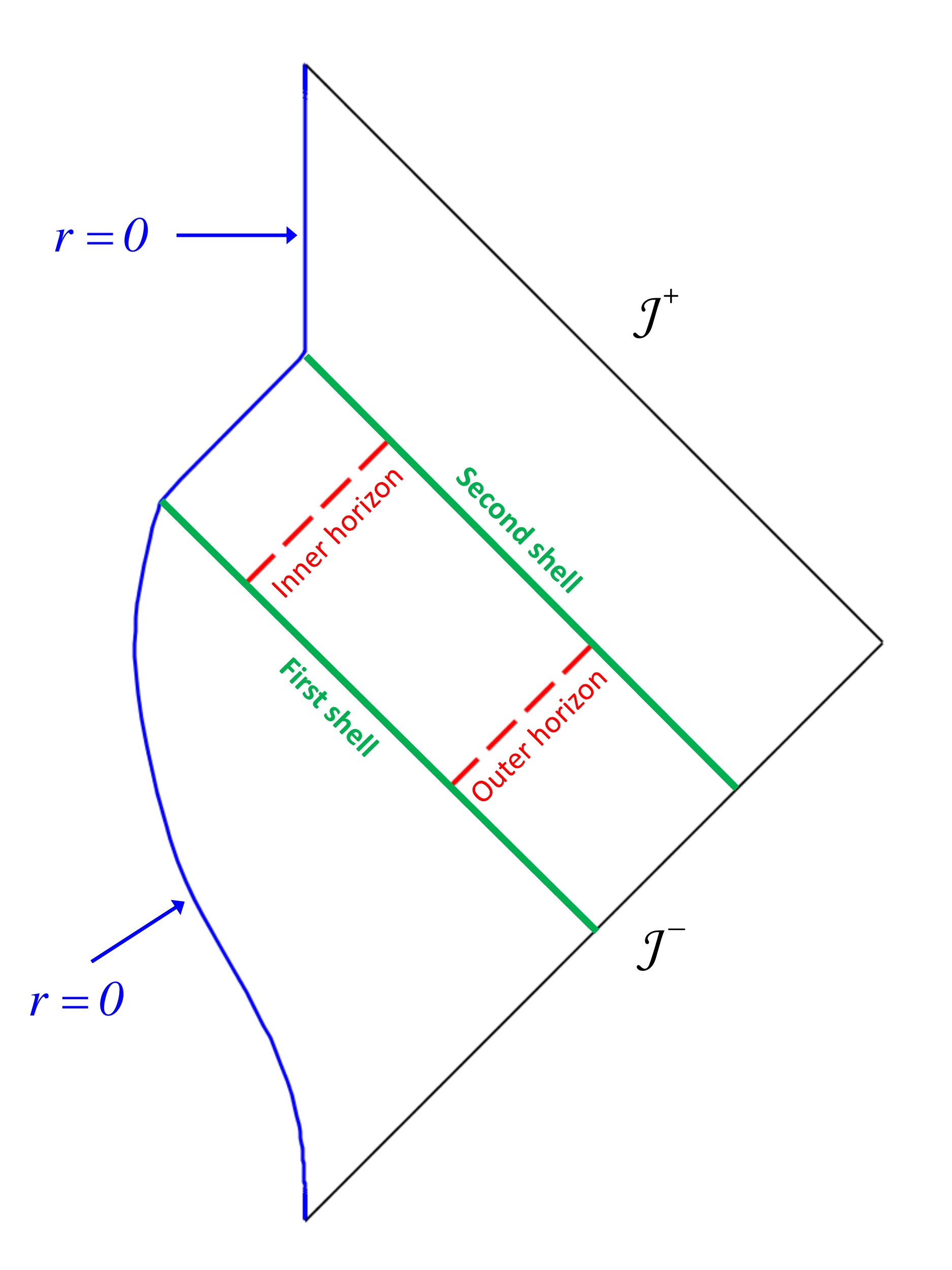}
  \caption{
  This  is a Carter-Penrose diagram for the modified $\alpha$-sandwich black hole with the parameters $p=3$, $q=3$, and $k=4, n=6$.
\label{PenroseA}}
\end{figure}

\subsection{Carter-Penrose diagram}

Fig.\,\ref{Uu_p4q8k4} shows the function $u_-(u_+)$ for the $\alpha$-sandwich black-hole with parameters $p=4, q=8$, and $k=4, n=6$. The difference with Fig.\,\ref{Uu_p4q8} appears from the red-shift factor $\alpha$ near the center of the black hole. For the chosen parameters of the $\alpha$-sandwich black-hole the outgoing null rays emitted from the center $r=0$ in the interval between the shells $v\in[0,q]$ do not have enough time to propagate far from the center. They are concentrated near $r=0$ and, when the second shell comes,  are released altogether from this narrow region and create an almost rectangular pulse of high frequency radiation propagating to infinity, the duration of this pulse is approximately $\alpha_0 q$. This radiation results from the domain, corresponding to $0< u_{-}<q$ (see an almost vertical part of the curve in Fig.\,\ref{Uu_p4q8k4}).

The Carter-Penrose diagram for the $\alpha$-sandwich black-hole with parameters $p=3, q=3$, and $k=4, n=6$  is shown in Fig.\,\ref{PenroseA}. We used the same coordinate transformations as in the Carter-Penrose diagram for standard sandwich black-hole Fig.\,\ref{Penrose}.  Consider null rays in the vicinity of the inner horizon. In the case of  the $\alpha$-sandwich black-hole, they cross the $r=0$ curve at larger angles than the standard case Fig.\,\ref{Penrose}. This reflects the lesser gain function and the lesser ultraviolet shift of quanta in the modified case. Note that in Fig.\,\ref{PenroseA} the $r=0$ curve between the shells looks very close to the null line, but, in fact, it is a timelike curve and its slope is a little bit bigger than 1 by about $\alpha_0=1/(1+p^{4})$.

\bigskip

\begin{figure}[tbp]
\centering
\includegraphics[width=7cm]{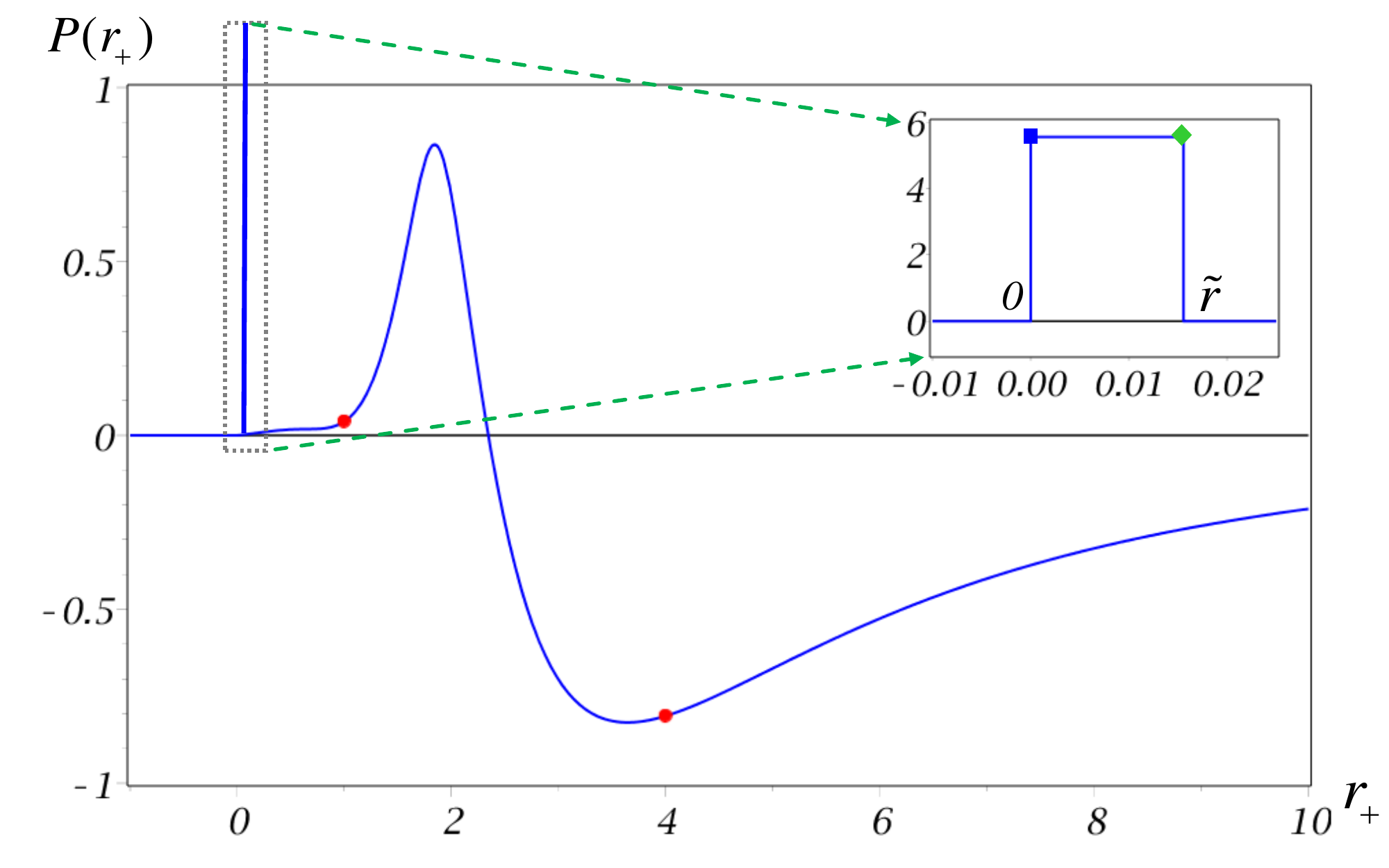}
  \caption{
  This plot illustrates the logarithm of the gain function $P(r_{+})=\ln\beta=\ln(du_{-}/du_{+})$ for the  $\alpha$-sandwich black hole with $p=4$, $q=8$, and $k=4, n=6$. Two red (circle) dots represent fluxes at the inner $r=1$  and outer $r=p$  horizons. The blue (square) marks $P$ at the point $r_+=0$. The green (diamond) dot marks $P$ at the point corresponding to $r_{-}=0$.
\label{P_p4q8k4}}
\end{figure}

\subsection{Gain function}

The logarithm of the gain function for the  $\alpha$-sandwich black-hole is shown in Fig.\,\ref{P_p4q8k4}. It has a peak at the point of the accumulation of rays. In comparison with the case of a standard black hole (see Fig.\,\ref{P_p4q8}) the peak of the gain function has lesser amplitude, looks smoother, and  is concentrated not at the inner horizon but mostly at some finite radius above it. The position of the peak can be found by solving numerically equations (\ref{FFp}) and (\ref{qF}). For example, this numerical solution for the chosen value of the parameters $p=4$ and $q=8$ gives the value $\rho=1.842$, which identically coincides with the position of the maximum of the peak of the logarithm of the gain function at the plot presented at Figure~\ref{P_p4q8k4}. We choose small value of the parameters $p$ and $q$ in order to be able to illustrate the behavior of the gain function in the total domain $r_+=\in [0,\infty)$. When $q$ becomes large, the height of the peak grows while its width becomes very narrow.

If we are interested in the case when both parameters $p$ and $q$ are large, one can obtain an approximate solution of the equations (\ref{FFp}) and (\ref{qF}). Thus we assume that $p$ is large. We also assume that the time of the existence of the sandwich black hole is larger than the Hawking evaporation time,  $q\ge p^3$.

For the  $\alpha$-sandwich black holes of the type \eq{dsalpha} the integral (\ref{qF}) can be taken exactly and be expressed in terms of elementary functions. Though the result is very cumbersome, one can  find its leading asymptotic for $q \ge p^3$. The result is
\be\begin{split}\label{kappa2q}
{q\over p^4} \approx &{1\over 6}\Big[6~\mathrm{arctanh}\left({1\over r}\right)+2\arctan{r}-5\pi\\
&+4\arctan(2r-\sqrt{3})+4\arctan(2r+\sqrt{3})\Big] \, .
\end{split}\ee
We checked numerically that \eq{kappa2q} is extremely accurate for all values of $q\ge p^3$. One can see that the dependence of $\rho$ on the lifespan of the black hole comes only via combination $q/p^4$. In the leading approximation the dependence of the integral on $\zeta(\rho)$ drops out. Any inaccuracy in the upper limit of the integral $\Delta \zeta(\rho)\sim p^{2/3}$ leads to a negligible correction $\Delta\rho\sim p^{-10/3}$.

This approximation can be also used to derive another fairly good approximation for the ${\cal F}(\rho(q))$, which is valid for all $p^3<q<\infty$ with an accuracy about $20\%$
\be\label{Fmax}
{\cal F}(\rho)=
\begin{cases}
-{1\over p^4}\left(1-{2 q\over 3p^4}\right)\left({7\over 2} {q\over p^4}\right)^{-8/7}\, ,&{p^3}<q < p^4\, ,\\
-{8\over p^4}\exp\left(-{2q\over p^4} -{5\pi\over 6}  \right)\, ,&q\ge p^4 \, .
\end{cases}
\ee
In the same approximation on gets
\be
\zeta(\rho)\approx 5^{1/6}p^{2/3}\hh
{\cal F}(\zeta(\rho))\approx -{1\over 6}\left( 5^{5/6}p^{1/3}-5\right) \, .
\ee
The maximum of the peak of the gain function can be estimated as follows
\be\label{betamax}
\beta_{max}\sim {|{\cal F}(\zeta(\rho))|\over |{\cal F}(\rho)|}\, .
\ee
One can see that for $p^3<q<p^4$ the maximum of the gain function grows as a power of $q$, while for very large $q\ge p^{4}$ it grows exponentially with $q$. The width of the peak of the gain function is proportional to ${\cal F}(\rho)$.

A new feature of the $\alpha$-sandwich model is the appearance of an almost rectangular pulse of the gain function at $r\in [0,\tilde{r}]$. Here $\tilde{r}$ is defined by the condition $r_{-}(\tilde{r})=0$. It means that the outgoing null ray emitted from the center $r=0$ at the moment of arrival of the first shell ($v=0$) reaches the second shell ($v=q$) at the radius $\tilde{r}$.

\bigskip

\begin{figure}[tbp]
\centering
\includegraphics[width=7cm]{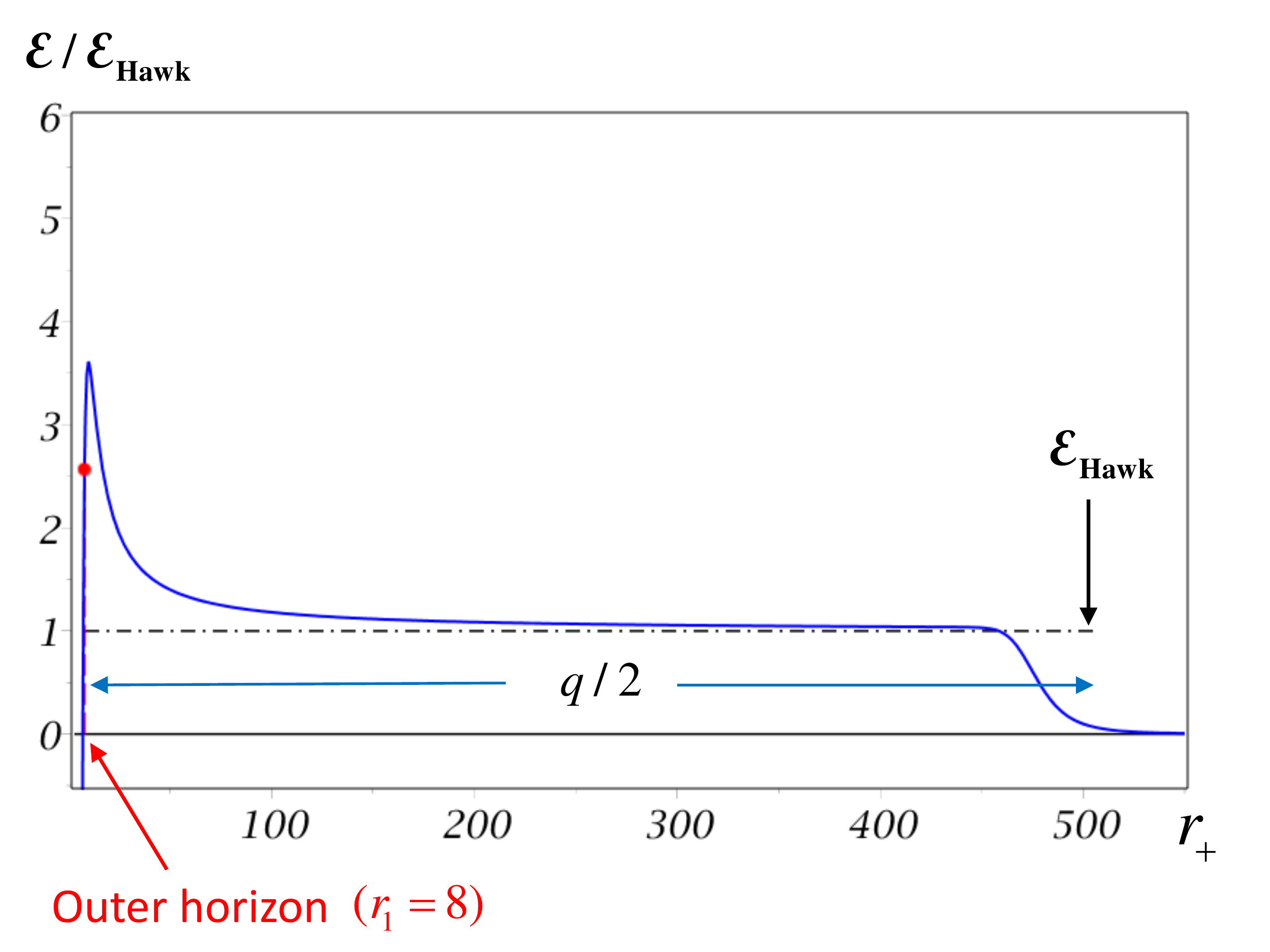}
  \caption{
  This plot depicts the flux of particles emitted from in the $\alpha$-sandwich black hole with $p=8$, $q=1000$, and $k=4, n=6$ at the moment $v=q$ in the range of radii from the outer horizon $r_1=8$ till infinity. The dash line denotes the asymptotic of the Hawking radiation in the limit of $q\to \infty$. At radii greater than $\approx q/2$ the flux vanishes. It corresponds to the coordinate $u_{-}\approx 0$. The tail of the Hawking radiation lasts $\approx q$, i.e., the lifespan of the black hole.
\label{Flux_p8q1000k4}}
\end{figure}

\subsection{Quantum radiation}

The Hawking radiation in the black hole exterior is practically the same as it is for the standard model with the same $p$ and $q$ parameters (see Fig.\,\ref{Flux_p8q1000k4}). In other words, the Hawking radiation practically does not depend on the internal structure of the black hole. The small difference in the flux is explained by the dependence of the surface gravity of the outer horizon on the fundamental constant $\ell$. For large $p$ this effect is negligible.

\begin{figure}[tbp]
\centering
\includegraphics[width=7cm]{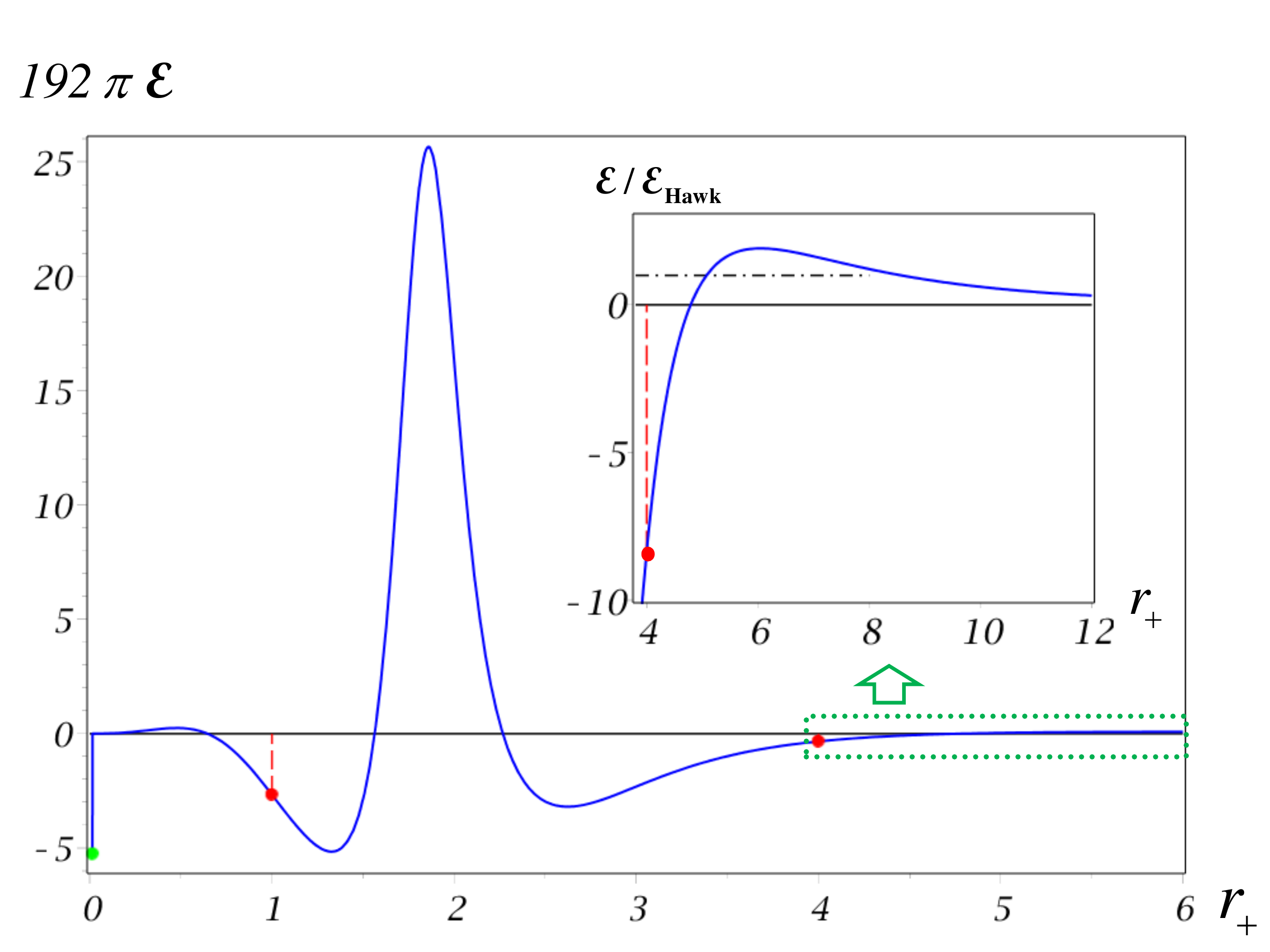}
  \caption{
  This plot depicts the flux of particles emitted by the $\alpha$-sandwich black hole with $p=4$, $q=8$, and $k=4, n=6$ at the moment $v=q$. The red dots on the plot correspond to the fluxes on the horizons at $r=1$ and $r=4$. We singled out the domain outside the outer horizon, rescaled it in the units of the Hawking flux, and put inside the figure. The green dot marks the point $r=\tilde{r}$.
  At radii $0< r<\tilde{r}$ the flux is negative and almost constant.
\label{Flux_p4q8k4}}
\end{figure}

Fig.\,\ref{Flux_p4q8k4} shows the energy flux from a  $\alpha$-sandwich black-hole. Here we chose a small mass black hole $p=4$ with the lifespan $q=8$. Though for these parameters the plateau of the Hawking radiation in the exterior of the black hole is less pronounced, the structure of the quantum radiation in its interior can be depicted in more details. The most important difference is the value of the amplitude of the peak-shaped radiation from the black hole interior. The position of the peak is shifted to a larger value of radius. Numerical calculations show that it practically coincides with the position of the gain function for the same parameters $p$ and $q$.
What is more important, for  $\alpha$-sandwich black-holes its height is greatly suppressed. If we increase the lifetime $q$ of the black hole, the peak becomes higher, its width decreases, and its position shifts closer to the inner horizon. At the same time the quantum radiation in the black hole exterior becomes closer and closer to the constant Hawking flux of duration $q$.

The amplitude of the peak radiation can be estimated using \eq{bb.2}. The maximum of quantum radiation at the moment $v=q$ comes out from the radius, which is very close to the radius  $\rho$ of the maximum of the gain function. For large $p$ and $q$ the value of the function
$B(r_{-})$ entering \eq{bb.2} can be roughly estimated at the point of a minimum of the function ${\cal F}$ (see \eq{Br}). Its value at this point is $B(\zeta(\rho))\sim 2.4\,p^{-2/3}$. Thus, the leading contribution to the flux density is
\be\label{Emax}
{\cal E}_\ins{max}\sim {B(\zeta(\rho))\over 192\pi {\cal F}^2(\rho)}.
\ee
In order to obtain the estimation for the rate of the energy flux at its maximum for large $p$ and $q\ge p^3$ it is sufficient to substitute the expression (\ref{Fmax}) for ${\cal F}^2(\rho)$ in this expression. As the result we get
\be\nonumber
192\pi {\cal E}_\ins{max}\sim
\begin{cases}
42~{p^{22/3}}{\left({q\over p^4}\right)^{16/7} \over\left(1-{2 q\over 3p^4}\right)^{2}},&{p^3}<q < p^4,\\
0.8~p^{22/3}\exp\left({4q \over p^4} \right),& q\ge p^4.
\end{cases}
\ee
In order to estimate the total energy emitted in this peak of radiation one has to know the width of the peak. The width of the peak of quantum radiation is basically the same as the width of the peak $\Delta r$ of the gain function. The latter can be estimated from \eq{betamax} and the condition $\Delta r\beta_\ins{max}\sim\int_{\tilde{r}}^{p} \beta(r_{+}) dr_{+}=\int_{0}^{p}dr_{-}=p$.
\be
[\mbox{width of the peak}]\sim\Delta r\sim p^{2/3}|{\cal F}(\rho)|\, .
\ee
Then the total energy flux for the black hole interior for $q\sim p^3$ can be estimated using \eq{Emax} as

\be
192\pi\Delta E_{\alpha}\sim 192\pi {\cal E}_\ins{max}\times \Delta r\sim |{\cal F}(\rho)|^{-1}\, .
\ee
Using \eq{Fmax} one can see that for a typical value of $q\sim p^3$ the asymptotic $|{\cal F}(\rho)|\sim p^{-20/7}$. Therefore, in this case
$192\pi\Delta E_{\alpha}\sim p^{-20/7}$.

The corresponding energy flux for the standard model for the same value of $q\sim p^3$ is
\be
\Delta E\sim \exp(p^3)\, .
\ee
This result means that inclusion of a properly chosen redshift function $\alpha$ allows one to supress the exponential outburst of the energy flux from the sandwich black hole interior. We have to stress that even after suppression the estimation of the total emitted energy $\sim p^{2.9}$ is larger than the initial mass of the black hole $\sim p$. Is this inconsistency a consequence of an adopted in this paper sandwich model with its unphysical switching-on and switching-off procedure? One can hope, that for more realistic evaporating black hole models with a proper dependence of mass on time, this inconsistency can be cured.

Note that the density of the energy fluxes can be either positive or negative. Negative flux densities may appear for a short period of time because of the quantum nature of the emitted radiation. Nevertheless, as we already mentioned, taking into account \eq{EEE} and asymptotic properties of the gain function that lead to the condition $P'\big|_{u_{+}=\pm\infty}=0$, one can show that total flux of quantum radiation is {\it always positive}.

\bigskip

\section{Discussion}

In the present paper we discuss quantum radiation from spherically symmetric sandwich-type nonsingular black holes. We calculate the quantum energy flux for two types black holes. In both cases the spacetime contains two null shell. The first shell, with a positive mass, creates the black hole. After some time this black hole is disrupted as a result of the collapse of the second shell with a negative mass. Two types of the sandwich black holes in question differ by the choice of the metric between the shells. In the first case, which we called a standard sandwich black hole, this metric coincides with Hayward solution. Such a model is characterized by two parameters: the mass and the duration parameters.  Using $2D$ approximation we calculated the energy flux of quantum massless scalar field in the background of this geometry. We demonstrated that after the formation of the black hole the calculated energy flux correctly reproduces the Hawking result. For a black hole of large mass and with large duration parameter the flux is determined by the surface gravity of the outer horizon and practically does not depend on the inner structure of the black hole. This Hawking regime of radiation continues during the interval of time of the black hole existence. This result clearly indicates that the outgoing quantum flux is formed near the outer horizon during its existence. The attempts to explain it as a result of the creation of the flux at the moment of formation of the black hole, which sometimes can be found in the literature, are wrong. A steady Hawking flux terminates at the time, when a signal from moment of intersection of the outer horizon by the second shell reaches an external observer.

Later the energy density of the radiation temporarily becomes negative. However, its amplitude remains rather small. This regime ends when a signal from the inner horizon reaches the observer. At this moment there exists a huge outburst of the radiation. The positive energy density in the peak is proportional to $\exp(2|\kappa_2|q \Delta V)$, where $\kappa_2$ is the (negative) surface density of the inner horizon. The width of the peak is $\sim \exp(-|\kappa_2| \Delta V)$, so that the total energy emitted during the outburst is exponentially large $\Delta E\sim \exp(|\kappa_2| \Delta V)$. A similar result can be expected  for any model of a black hole with the  inner horizon with $|\kappa_1|\sim \ell^{-1}$. For the large duration $\Delta V \gg M$ the emitted energy from this domain is much larger that the black hole mass $M$. Thus such a model is not self-consistent.

Hence, the back-reaction of the quanta propagating near the inner horizon on the geometry should greatly modify the geometry in this domain.
One of the options is that after the back-reaction is included, the surface gravity of the inner horizon becomes small. In order to demonstrate how the exponentially large radiation from the inner horizon can be suppressed, we considered a so called ``modified model'', proposed in \cite{Frolov:2016pav,DeLorenzo:2014pta,DeLorenzo:2015taa}. This model includes a non-trivial redshift function $\alpha(r)$. We normalize this factor so that at infinity it always has the value $1$. In such a case the value $\alpha_0$ of this parameter at the center characterizes the redshift of a photon, propagating from infinity to the center. Important new feature of a $\alpha$-sandwich black hole is that the surface gravity of the inner horizon depends on the choice of $\alpha$ and can be made small.

We demonstrated that the gain function plays an important role in understanding features of the quantum radiation from the black hole interior. As we showed the peak of the gain function is a result of the effect of the accumulation of the outgoing null rays at some point in the black hole interior. The position of this point depends on the time $q$ of the existence of the sandwich black hole. This effect is generic for nonsingular black holes. The reason is that in the vicinity of the center the gravity is not attractive, but repulsive. As a result the velocity of the falling down null rays inside the apparent horizon is slowed down and the radial density of a beam of photon increases. We obtained equations which determine a position of the peak of the gain function for a given parameter $q$. Numerical computations of the quantum energy flux from the interior of the black hole demonstrated that an amplitude and a position of the peak of the energy flux are strongly correlated with similar characteristics of the peak of the gain function. We used this result to obtain estimation of the energy flux from the black hole interior for large mass $\sim p$ and $q\ge p^3$. These results are applicable for the both, standard and $\alpha$ sandwich black holes. The main result is that the exponential peak of the energy outburst, which is characteristic for the standard sandwich black holes is strongly suppressed for the $\alpha$-sandwich model. For a concrete $\alpha$-sandwich model, which we considered in the paper, the total quantum energy, radiated from the interior of a black hole with $q\sim p^3$ is proportional to $\sim p^{2.9}$. It seams that the exponential dependence of the energy emitted in the standard sandwich model is a robust prediction of models with a trivial redshift function $\alpha=1$. The power in mass energy release in the $\alpha$-models is still larger than the initial mass of the black hole $\sim p$. However, it might be that this result is a manifestation of the adopted model roughness. In order to check this, one has to repeat the calculations, presented in this paper, to a more ``realistic'' evaporating nonsingular black model. Let us mention that an independent interesting approach to the self-consistency problem are attempts to construct a regular black hole model in the framework of a $2D$ dilaton gravity, that include both quantum radiation and its back-reaction (see e.g. \cite{Trodden:1993dm, Easson:2002tg, Taves:2014laa, Kunstatter:2015vxa}.
Anyway, search for (quantum) consistent nonsingular models of black holes is an interesting open problem which waits for its solution.


\section*{Acknowledgments}

The authors thank the Natural Sciences and Engineering Research Council of Canada and the Killam Trust for their financial support.


\begin{thebibliography}{62}%
\makeatletter
\providecommand \@ifxundefined [1]{%
 \@ifx{#1\undefined}
}%
\providecommand \@ifnum [1]{%
 \ifnum #1\expandafter \@firstoftwo
 \else \expandafter \@secondoftwo
 \fi
}%
\providecommand \@ifx [1]{%
 \ifx #1\expandafter \@firstoftwo
 \else \expandafter \@secondoftwo
 \fi
}%
\providecommand \natexlab [1]{#1}%
\providecommand \enquote  [1]{``#1''}%
\providecommand \bibnamefont  [1]{#1}%
\providecommand \bibfnamefont [1]{#1}%
\providecommand \citenamefont [1]{#1}%
\providecommand \href@noop [0]{\@secondoftwo}%
\providecommand \href [0]{\begingroup \@sanitize@url \@href}%
\providecommand \@href[1]{\@@startlink{#1}\@@href}%
\providecommand \@@href[1]{\endgroup#1\@@endlink}%
\providecommand \@sanitize@url [0]{\catcode `\\12\catcode `\$12\catcode
  `\&12\catcode `\#12\catcode `\^12\catcode `\_12\catcode `\%12\relax}%
\providecommand \@@startlink[1]{}%
\providecommand \@@endlink[0]{}%
\providecommand \url  [0]{\begingroup\@sanitize@url \@url }%
\providecommand \@url [1]{\endgroup\@href {#1}{\urlprefix }}%
\providecommand \urlprefix  [0]{URL }%
\providecommand \Eprint [0]{\href }%
\providecommand \doibase [0]{http://dx.doi.org/}%
\providecommand \selectlanguage [0]{\@gobble}%
\providecommand \bibinfo  [0]{\@secondoftwo}%
\providecommand \bibfield  [0]{\@secondoftwo}%
\providecommand \translation [1]{[#1]}%
\providecommand \BibitemOpen [0]{}%
\providecommand \bibitemStop [0]{}%
\providecommand \bibitemNoStop [0]{.\EOS\space}%
\providecommand \EOS [0]{\spacefactor3000\relax}%
\providecommand \BibitemShut  [1]{\csname bibitem#1\endcsname}%
\let\auto@bib@innerbib\@empty
\bibitem [{\citenamefont {Tomboulis}(1997)}]{Tomboulis:1997gg}%
  \BibitemOpen
  \bibfield  {author} {\bibinfo {author} {\bibfnamefont {E.}~\bibnamefont
  {Tomboulis}},\ }\bibfield  {title} {\emph {\enquote {\bibinfo {title}
  {{Superrenormalizable gauge and gravitational theories}},}\ }}\href@noop {}
  {\  (\bibinfo {year} {1997})},\ \Eprint {http://arxiv.org/abs/hep-th/9702146}
  {arXiv:hep-th/9702146 [hep-th]} \BibitemShut {NoStop}%
\bibitem [{\citenamefont {Biswas}\ \emph {et~al.}(2012)\citenamefont {Biswas},
  \citenamefont {Gerwick}, \citenamefont {Koivisto},\ and\ \citenamefont
  {Mazumdar}}]{Biswas:2011ar}%
  \BibitemOpen
  \bibfield  {author} {\bibinfo {author} {\bibfnamefont {T.}~\bibnamefont
  {Biswas}}, \bibinfo {author} {\bibfnamefont {E.}~\bibnamefont {Gerwick}},
  \bibinfo {author} {\bibfnamefont {T.}~\bibnamefont {Koivisto}}, \ and\
  \bibinfo {author} {\bibfnamefont {A.}~\bibnamefont {Mazumdar}},\ }\bibfield
  {title} {\emph {\enquote {\bibinfo {title} {{Towards singularity and ghost
  free theories of gravity}},}\ }}\href {\doibase
  10.1103/PhysRevLett.108.031101} {\bibfield  {journal} {\bibinfo  {journal}
  {Phys. Rev. Lett.}\ }\textbf {\bibinfo {volume} {108}},\ \bibinfo {pages}
  {031101} (\bibinfo {year} {2012})},\ \Eprint {http://arxiv.org/abs/1110.5249}
  {arXiv:1110.5249 [gr-qc]} \BibitemShut {NoStop}%
\bibitem [{\citenamefont {Modesto}(2012{\natexlab{a}})}]{Modesto:2011kw}%
  \BibitemOpen
  \bibfield  {author} {\bibinfo {author} {\bibfnamefont {L.}~\bibnamefont
  {Modesto}},\ }\bibfield  {title} {\emph {\enquote {\bibinfo {title}
  {{Super-renormalizable Quantum Gravity}},}\ }}\href {\doibase
  10.1103/PhysRevD.86.044005} {\bibfield  {journal} {\bibinfo  {journal}
  {Phys.Rev.}\ }\textbf {\bibinfo {volume} {D86}},\ \bibinfo {pages} {044005}
  (\bibinfo {year} {2012}{\natexlab{a}})},\ \Eprint
  {http://arxiv.org/abs/1107.2403} {arXiv:1107.2403 [hep-th]} \BibitemShut
  {NoStop}%
\bibitem [{\citenamefont {Modesto}(2012{\natexlab{b}})}]{Modesto:2012ys}%
  \BibitemOpen
  \bibfield  {author} {\bibinfo {author} {\bibfnamefont {L.}~\bibnamefont
  {Modesto}},\ }\bibfield  {title} {\emph {\enquote {\bibinfo {title}
  {{Super-Renormalizable Multidimensional Gravity: Theory and Applications}},}\
  }}\href@noop {} {\bibfield  {journal} {\bibinfo  {journal} {Astron.Rev.}\
  }\textbf {\bibinfo {volume} {8}},\ \bibinfo {pages} {4} (\bibinfo {year}
  {2012}{\natexlab{b}})},\ \Eprint {http://arxiv.org/abs/1202.3151}
  {arXiv:1202.3151 [hep-th]} \BibitemShut {NoStop}%
\bibitem [{\citenamefont {Biswas}\ \emph {et~al.}(2014)\citenamefont {Biswas},
  \citenamefont {Conroy}, \citenamefont {Koshelev},\ and\ \citenamefont
  {Mazumdar}}]{Biswas:2013cha}%
  \BibitemOpen
  \bibfield  {author} {\bibinfo {author} {\bibfnamefont {T.}~\bibnamefont
  {Biswas}}, \bibinfo {author} {\bibfnamefont {A.}~\bibnamefont {Conroy}},
  \bibinfo {author} {\bibfnamefont {A.~S.}\ \bibnamefont {Koshelev}}, \ and\
  \bibinfo {author} {\bibfnamefont {A.}~\bibnamefont {Mazumdar}},\ }\bibfield
  {title} {\emph {\enquote {\bibinfo {title} {{Generalized ghost-free quadratic
  curvature gravity}},}\ }}\href {\doibase 10.1088/0264-9381/31/1/015022,
  10.1088/0264-9381/31/15/159501} {\bibfield  {journal} {\bibinfo  {journal}
  {Class.Quant.Grav.}\ }\textbf {\bibinfo {volume} {31}},\ \bibinfo {pages}
  {015022} (\bibinfo {year} {2014})},\ \Eprint {http://arxiv.org/abs/1308.2319}
  {arXiv:1308.2319 [hep-th]} \BibitemShut {NoStop}%
\bibitem [{\citenamefont {Biswas}\ \emph {et~al.}(2013)\citenamefont {Biswas},
  \citenamefont {Koivisto},\ and\ \citenamefont {Mazumdar}}]{Biswas:2013kla}%
  \BibitemOpen
  \bibfield  {author} {\bibinfo {author} {\bibfnamefont {T.}~\bibnamefont
  {Biswas}}, \bibinfo {author} {\bibfnamefont {T.}~\bibnamefont {Koivisto}}, \
  and\ \bibinfo {author} {\bibfnamefont {A.}~\bibnamefont {Mazumdar}},\ }in\
  \href {http://inspirehep.net/record/1217538/files/arXiv:1302.0532.pdf} {\emph
  {\bibinfo {booktitle} {{Proceedings, Barcelona Postgrad Encounters on
  Fundamental Physics}}}}\ (\bibinfo {year} {2013})\ pp.\ \bibinfo {pages}
  {13--24},\ \Eprint {http://arxiv.org/abs/1302.0532} {arXiv:1302.0532 [gr-qc]}
  \BibitemShut {NoStop}%
\bibitem [{\citenamefont {Modesto}\ and\ \citenamefont
  {Rachwal}(2014)}]{Modesto:2014lga}%
  \BibitemOpen
  \bibfield  {author} {\bibinfo {author} {\bibfnamefont {L.}~\bibnamefont
  {Modesto}}\ and\ \bibinfo {author} {\bibfnamefont {L.}~\bibnamefont
  {Rachwal}},\ }\bibfield  {title} {\emph {\enquote {\bibinfo {title}
  {{Super-renormalizable and finite gravitational theories}},}\ }}\href
  {\doibase 10.1016/j.nuclphysb.2014.10.015} {\bibfield  {journal} {\bibinfo
  {journal} {Nucl.Phys.}\ }\textbf {\bibinfo {volume} {B889}},\ \bibinfo
  {pages} {228} (\bibinfo {year} {2014})},\ \Eprint
  {http://arxiv.org/abs/1407.8036} {arXiv:1407.8036 [hep-th]} \BibitemShut
  {NoStop}%
\bibitem [{\citenamefont {Talaganis}\ \emph {et~al.}(2015)\citenamefont
  {Talaganis}, \citenamefont {Biswas},\ and\ \citenamefont
  {Mazumdar}}]{Talaganis:2014ida}%
  \BibitemOpen
  \bibfield  {author} {\bibinfo {author} {\bibfnamefont {S.}~\bibnamefont
  {Talaganis}}, \bibinfo {author} {\bibfnamefont {T.}~\bibnamefont {Biswas}}, \
  and\ \bibinfo {author} {\bibfnamefont {A.}~\bibnamefont {Mazumdar}},\
  }\bibfield  {title} {\emph {\enquote {\bibinfo {title} {{Towards
  understanding the ultraviolet behavior of quantum loops in
  infinite-derivative theories of gravity}},}\ }}\href {\doibase
  10.1088/0264-9381/32/21/215017} {\bibfield  {journal} {\bibinfo  {journal}
  {Class. Quant. Grav.}\ }\textbf {\bibinfo {volume} {32}},\ \bibinfo {pages}
  {215017} (\bibinfo {year} {2015})},\ \Eprint {http://arxiv.org/abs/1412.3467}
  {arXiv:1412.3467 [hep-th]} \BibitemShut {NoStop}%
\bibitem [{\citenamefont {Tomboulis}(2015{\natexlab{a}})}]{Tomboulis:2015gfa}%
  \BibitemOpen
  \bibfield  {author} {\bibinfo {author} {\bibfnamefont {E.~T.}\ \bibnamefont
  {Tomboulis}},\ }\bibfield  {title} {\emph {\enquote {\bibinfo {title}
  {{Nonlocal and quasi-local field theories}},}\ }}\href@noop {} {\  (\bibinfo
  {year} {2015}{\natexlab{a}})},\ \Eprint {http://arxiv.org/abs/1507.00981}
  {arXiv:1507.00981 [hep-th]} \BibitemShut {NoStop}%
\bibitem [{\citenamefont {Tomboulis}(2015{\natexlab{b}})}]{Tomboulis:2015esa}%
  \BibitemOpen
  \bibfield  {author} {\bibinfo {author} {\bibfnamefont {E.~T.}\ \bibnamefont
  {Tomboulis}},\ }\bibfield  {title} {\emph {\enquote {\bibinfo {title}
  {{Renormalization and unitarity in higher derivative and nonlocal gravity
  theories}},}\ }}\href {\doibase 10.1142/S0217732315400052} {\bibfield
  {journal} {\bibinfo  {journal} {Mod. Phys. Lett.}\ }\textbf {\bibinfo
  {volume} {A30}},\ \bibinfo {pages} {1540005} (\bibinfo {year}
  {2015}{\natexlab{b}})}\BibitemShut {NoStop}%
\bibitem [{\citenamefont {Nicolini}\ \emph {et~al.}(2006)\citenamefont
  {Nicolini}, \citenamefont {Smailagic},\ and\ \citenamefont
  {Spallucci}}]{Nicolini:2005vd}%
  \BibitemOpen
  \bibfield  {author} {\bibinfo {author} {\bibfnamefont {P.}~\bibnamefont
  {Nicolini}}, \bibinfo {author} {\bibfnamefont {A.}~\bibnamefont {Smailagic}},
  \ and\ \bibinfo {author} {\bibfnamefont {E.}~\bibnamefont {Spallucci}},\
  }\bibfield  {title} {\emph {\enquote {\bibinfo {title} {{Noncommutative
  geometry inspired Schwarzschild black hole}},}\ }}\href {\doibase
  10.1016/j.physletb.2005.11.004} {\bibfield  {journal} {\bibinfo  {journal}
  {Phys. Lett.}\ }\textbf {\bibinfo {volume} {B632}},\ \bibinfo {pages} {547}
  (\bibinfo {year} {2006})},\ \Eprint {http://arxiv.org/abs/gr-qc/0510112}
  {arXiv:gr-qc/0510112 [gr-qc]} \BibitemShut {NoStop}%
\bibitem [{\citenamefont {Spallucci}\ \emph {et~al.}(2006)\citenamefont
  {Spallucci}, \citenamefont {Smailagic},\ and\ \citenamefont
  {Nicolini}}]{Spallucci:2006zj}%
  \BibitemOpen
  \bibfield  {author} {\bibinfo {author} {\bibfnamefont {E.}~\bibnamefont
  {Spallucci}}, \bibinfo {author} {\bibfnamefont {A.}~\bibnamefont
  {Smailagic}}, \ and\ \bibinfo {author} {\bibfnamefont {P.}~\bibnamefont
  {Nicolini}},\ }\bibfield  {title} {\emph {\enquote {\bibinfo {title} {{Trace
  Anomaly in Quantum Spacetime Manifold}},}\ }}\href {\doibase
  10.1103/PhysRevD.73.084004} {\bibfield  {journal} {\bibinfo  {journal} {Phys.
  Rev.}\ }\textbf {\bibinfo {volume} {D73}},\ \bibinfo {pages} {084004}
  (\bibinfo {year} {2006})},\ \Eprint {http://arxiv.org/abs/hep-th/0604094}
  {arXiv:hep-th/0604094 [hep-th]} \BibitemShut {NoStop}%
\bibitem [{\citenamefont {Nicolini}(2005)}]{Nicolini:2005zi}%
  \BibitemOpen
  \bibfield  {author} {\bibinfo {author} {\bibfnamefont {P.}~\bibnamefont
  {Nicolini}},\ }\bibfield  {title} {\emph {\enquote {\bibinfo {title} {{A
  Model of radiating black hole in noncommutative geometry}},}\ }}\href
  {\doibase 10.1088/0305-4470/38/39/L02} {\bibfield  {journal} {\bibinfo
  {journal} {J. Phys.}\ }\textbf {\bibinfo {volume} {A38}},\ \bibinfo {pages}
  {L631} (\bibinfo {year} {2005})},\ \Eprint
  {http://arxiv.org/abs/hep-th/0507266} {arXiv:hep-th/0507266 [hep-th]}
  \BibitemShut {NoStop}%
\bibitem [{\citenamefont {Biswas}\ \emph {et~al.}(2010)\citenamefont {Biswas},
  \citenamefont {Koivisto},\ and\ \citenamefont {Mazumdar}}]{Biswas:2010zk}%
  \BibitemOpen
  \bibfield  {author} {\bibinfo {author} {\bibfnamefont {T.}~\bibnamefont
  {Biswas}}, \bibinfo {author} {\bibfnamefont {T.}~\bibnamefont {Koivisto}}, \
  and\ \bibinfo {author} {\bibfnamefont {A.}~\bibnamefont {Mazumdar}},\
  }\bibfield  {title} {\emph {\enquote {\bibinfo {title} {{Towards a resolution
  of the cosmological singularity in non-local higher derivative theories of
  gravity}},}\ }}\href {\doibase 10.1088/1475-7516/2010/11/008} {\bibfield
  {journal} {\bibinfo  {journal} {JCAP}\ }\textbf {\bibinfo {volume} {1011}},\
  \bibinfo {pages} {008} (\bibinfo {year} {2010})},\ \Eprint
  {http://arxiv.org/abs/1005.0590} {arXiv:1005.0590 [hep-th]} \BibitemShut
  {NoStop}%
\bibitem [{\citenamefont {Modesto}\ \emph {et~al.}(2011)\citenamefont
  {Modesto}, \citenamefont {Moffat},\ and\ \citenamefont
  {Nicolini}}]{Modesto:2010uh}%
  \BibitemOpen
  \bibfield  {author} {\bibinfo {author} {\bibfnamefont {L.}~\bibnamefont
  {Modesto}}, \bibinfo {author} {\bibfnamefont {J.~W.}\ \bibnamefont {Moffat}},
  \ and\ \bibinfo {author} {\bibfnamefont {P.}~\bibnamefont {Nicolini}},\
  }\bibfield  {title} {\emph {\enquote {\bibinfo {title} {{Black holes in an
  ultraviolet complete quantum gravity}},}\ }}\href {\doibase
  10.1016/j.physletb.2010.11.046} {\bibfield  {journal} {\bibinfo  {journal}
  {Phys.Lett.}\ }\textbf {\bibinfo {volume} {B695}},\ \bibinfo {pages} {397}
  (\bibinfo {year} {2011})},\ \Eprint {http://arxiv.org/abs/1010.0680}
  {arXiv:1010.0680 [gr-qc]} \BibitemShut {NoStop}%
\bibitem [{\citenamefont {Hossenfelder}\ \emph {et~al.}(2010)\citenamefont
  {Hossenfelder}, \citenamefont {Modesto},\ and\ \citenamefont
  {Premont-Schwarz}}]{Hossenfelder:2009fc}%
  \BibitemOpen
  \bibfield  {author} {\bibinfo {author} {\bibfnamefont {S.}~\bibnamefont
  {Hossenfelder}}, \bibinfo {author} {\bibfnamefont {L.}~\bibnamefont
  {Modesto}}, \ and\ \bibinfo {author} {\bibfnamefont {I.}~\bibnamefont
  {Premont-Schwarz}},\ }\bibfield  {title} {\emph {\enquote {\bibinfo {title}
  {{A Model for non-singular black hole collapse and evaporation}},}\ }}\href
  {\doibase 10.1103/PhysRevD.81.044036} {\bibfield  {journal} {\bibinfo
  {journal} {Phys. Rev.}\ }\textbf {\bibinfo {volume} {D81}},\ \bibinfo {pages}
  {044036} (\bibinfo {year} {2010})},\ \Eprint {http://arxiv.org/abs/0912.1823}
  {arXiv:0912.1823 [gr-qc]} \BibitemShut {NoStop}%
\bibitem [{\citenamefont {Calcagni}\ \emph {et~al.}(2014)\citenamefont
  {Calcagni}, \citenamefont {Modesto},\ and\ \citenamefont
  {Nicolini}}]{Calcagni:2013vra}%
  \BibitemOpen
  \bibfield  {author} {\bibinfo {author} {\bibfnamefont {G.}~\bibnamefont
  {Calcagni}}, \bibinfo {author} {\bibfnamefont {L.}~\bibnamefont {Modesto}}, \
  and\ \bibinfo {author} {\bibfnamefont {P.}~\bibnamefont {Nicolini}},\
  }\bibfield  {title} {\emph {\enquote {\bibinfo {title} {{Super-accelerating
  bouncing cosmology in asymptotically-free non-local gravity}},}\ }}\href
  {\doibase 10.1140/epjc/s10052-014-2999-8} {\bibfield  {journal} {\bibinfo
  {journal} {Eur. Phys. J.}\ }\textbf {\bibinfo {volume} {C74}},\ \bibinfo
  {pages} {2999} (\bibinfo {year} {2014})},\ \Eprint
  {http://arxiv.org/abs/1306.5332} {arXiv:1306.5332 [gr-qc]} \BibitemShut
  {NoStop}%
\bibitem [{\citenamefont {Zhang}\ \emph {et~al.}(2015)\citenamefont {Zhang},
  \citenamefont {Zhu}, \citenamefont {Modesto},\ and\ \citenamefont
  {Bambi}}]{Zhang:2014bea}%
  \BibitemOpen
  \bibfield  {author} {\bibinfo {author} {\bibfnamefont {Y.}~\bibnamefont
  {Zhang}}, \bibinfo {author} {\bibfnamefont {Y.}~\bibnamefont {Zhu}}, \bibinfo
  {author} {\bibfnamefont {L.}~\bibnamefont {Modesto}}, \ and\ \bibinfo
  {author} {\bibfnamefont {C.}~\bibnamefont {Bambi}},\ }\bibfield  {title}
  {\emph {\enquote {\bibinfo {title} {{Can static regular black holes form from
  gravitational collapse?}}}\ }}\href {\doibase 10.1140/epjc/s10052-015-3311-2}
  {\bibfield  {journal} {\bibinfo  {journal} {Eur. Phys. J.}\ }\textbf
  {\bibinfo {volume} {C75}},\ \bibinfo {pages} {96} (\bibinfo {year} {2015})},\
  \Eprint {http://arxiv.org/abs/1404.4770} {arXiv:1404.4770 [gr-qc]}
  \BibitemShut {NoStop}%
\bibitem [{\citenamefont {Conroy}\ \emph {et~al.}(2015)\citenamefont {Conroy},
  \citenamefont {Mazumdar},\ and\ \citenamefont {Teimouri}}]{Conroy:2015wfa}%
  \BibitemOpen
  \bibfield  {author} {\bibinfo {author} {\bibfnamefont {A.}~\bibnamefont
  {Conroy}}, \bibinfo {author} {\bibfnamefont {A.}~\bibnamefont {Mazumdar}}, \
  and\ \bibinfo {author} {\bibfnamefont {A.}~\bibnamefont {Teimouri}},\
  }\bibfield  {title} {\emph {\enquote {\bibinfo {title} {{Wald Entropy for
  Ghost-Free, Infinite Derivative Theories of Gravity}},}\ }}\href {\doibase
  10.1103/PhysRevLett.114.201101} {\bibfield  {journal} {\bibinfo  {journal}
  {Phys. Rev. Lett.}\ }\textbf {\bibinfo {volume} {114}},\ \bibinfo {pages}
  {201101} (\bibinfo {year} {2015})},\ \Eprint
  {http://arxiv.org/abs/1503.05568} {arXiv:1503.05568 [hep-th]} \BibitemShut
  {NoStop}%
\bibitem [{\citenamefont {Frolov}\ \emph {et~al.}(2015)\citenamefont {Frolov},
  \citenamefont {Zelnikov},\ and\ \citenamefont
  {de~Paula~Netto}}]{Frolov:2015bia}%
  \BibitemOpen
  \bibfield  {author} {\bibinfo {author} {\bibfnamefont {V.~P.}\ \bibnamefont
  {Frolov}}, \bibinfo {author} {\bibfnamefont {A.}~\bibnamefont {Zelnikov}}, \
  and\ \bibinfo {author} {\bibfnamefont {T.}~\bibnamefont {de~Paula~Netto}},\
  }\bibfield  {title} {\emph {\enquote {\bibinfo {title} {{Spherical collapse
  of small masses in the ghost-free gravity}},}\ }}\href {\doibase
  10.1007/JHEP06(2015)107} {\bibfield  {journal} {\bibinfo  {journal} {JHEP}\
  }\textbf {\bibinfo {volume} {06}},\ \bibinfo {pages} {107} (\bibinfo {year}
  {2015})},\ \Eprint {http://arxiv.org/abs/1504.00412} {arXiv:1504.00412
  [hep-th]} \BibitemShut {NoStop}%
\bibitem [{\citenamefont {Frolov}(2015)}]{Frolov:2015bta}%
  \BibitemOpen
  \bibfield  {author} {\bibinfo {author} {\bibfnamefont {V.~P.}\ \bibnamefont
  {Frolov}},\ }\bibfield  {title} {\emph {\enquote {\bibinfo {title} {{Mass-gap
  for black hole formation in higher derivative and ghost free gravity}},}\
  }}\href {\doibase 10.1103/PhysRevLett.115.051102} {\bibfield  {journal}
  {\bibinfo  {journal} {Phys. Rev. Lett.}\ }\textbf {\bibinfo {volume} {115}},\
  \bibinfo {pages} {051102} (\bibinfo {year} {2015})},\ \Eprint
  {http://arxiv.org/abs/1505.00492} {arXiv:1505.00492 [hep-th]} \BibitemShut
  {NoStop}%
\bibitem [{\citenamefont {Frolov}\ and\ \citenamefont
  {Zelnikov}(2015)}]{Frolov:2015usa}%
  \BibitemOpen
  \bibfield  {author} {\bibinfo {author} {\bibfnamefont {V.~P.}\ \bibnamefont
  {Frolov}}\ and\ \bibinfo {author} {\bibfnamefont {A.}~\bibnamefont
  {Zelnikov}},\ }\bibfield  {title} {\emph {\enquote {\bibinfo {title}
  {{Head-on collision of ultra-relativistic particles in ghost-free theories of
  gravity}},}\ }}\href@noop {} {\  (\bibinfo {year} {2015})},\ \Eprint
  {http://arxiv.org/abs/1509.03336} {arXiv:1509.03336 [hep-th]} \BibitemShut
  {NoStop}%
\bibitem [{\citenamefont {Li}\ \emph {et~al.}(2015)\citenamefont {Li},
  \citenamefont {Modesto},\ and\ \citenamefont {Rachwal}}]{Li:2015bqa}%
  \BibitemOpen
  \bibfield  {author} {\bibinfo {author} {\bibfnamefont {Y.-D.}\ \bibnamefont
  {Li}}, \bibinfo {author} {\bibfnamefont {L.}~\bibnamefont {Modesto}}, \ and\
  \bibinfo {author} {\bibfnamefont {L.}~\bibnamefont {Rachwal}},\ }\bibfield
  {title} {\emph {\enquote {\bibinfo {title} {{Exact solutions and spacetime
  singularities in nonlocal gravity}},}\ }}\href@noop {} {\  (\bibinfo {year}
  {2015})},\ \Eprint {http://arxiv.org/abs/1506.08619} {arXiv:1506.08619
  [hep-th]} \BibitemShut {NoStop}%
\bibitem [{\citenamefont {Bambi}\ \emph
  {et~al.}(2016{\natexlab{a}})\citenamefont {Bambi}, \citenamefont
  {Malafarina},\ and\ \citenamefont {Modesto}}]{Bambi:2016uda}%
  \BibitemOpen
  \bibfield  {author} {\bibinfo {author} {\bibfnamefont {C.}~\bibnamefont
  {Bambi}}, \bibinfo {author} {\bibfnamefont {D.}~\bibnamefont {Malafarina}}, \
  and\ \bibinfo {author} {\bibfnamefont {L.}~\bibnamefont {Modesto}},\
  }\bibfield  {title} {\emph {\enquote {\bibinfo {title} {{Black supernovae and
  black holes in non-local gravity}},}\ }}\href {\doibase
  10.1007/JHEP04(2016)147} {\bibfield  {journal} {\bibinfo  {journal} {JHEP}\
  }\textbf {\bibinfo {volume} {04}},\ \bibinfo {pages} {147} (\bibinfo {year}
  {2016}{\natexlab{a}})},\ \Eprint {http://arxiv.org/abs/1603.09592}
  {arXiv:1603.09592 [gr-qc]} \BibitemShut {NoStop}%
\bibitem [{\citenamefont {Bambi}\ \emph
  {et~al.}(2016{\natexlab{b}})\citenamefont {Bambi}, \citenamefont {Modesto},
  \citenamefont {Porey},\ and\ \citenamefont {Rachwal}}]{Bambi:2016yne}%
  \BibitemOpen
  \bibfield  {author} {\bibinfo {author} {\bibfnamefont {C.}~\bibnamefont
  {Bambi}}, \bibinfo {author} {\bibfnamefont {L.}~\bibnamefont {Modesto}},
  \bibinfo {author} {\bibfnamefont {S.}~\bibnamefont {Porey}}, \ and\ \bibinfo
  {author} {\bibfnamefont {L.}~\bibnamefont {Rachwal}},\ }\bibfield  {title}
  {\emph {\enquote {\bibinfo {title} {{Black hole evaporation in conformal
  gravity}},}\ }}\href@noop {} {\  (\bibinfo {year} {2016}{\natexlab{b}})},\
  \Eprint {http://arxiv.org/abs/1611.05582} {arXiv:1611.05582 [gr-qc]}
  \BibitemShut {NoStop}%
\bibitem [{\citenamefont {Bambi}\ \emph
  {et~al.}(2016{\natexlab{c}})\citenamefont {Bambi}, \citenamefont {Modesto},\
  and\ \citenamefont {Wang}}]{Bambi:2016wmo}%
  \BibitemOpen
  \bibfield  {author} {\bibinfo {author} {\bibfnamefont {C.}~\bibnamefont
  {Bambi}}, \bibinfo {author} {\bibfnamefont {L.}~\bibnamefont {Modesto}}, \
  and\ \bibinfo {author} {\bibfnamefont {Y.}~\bibnamefont {Wang}},\ }\bibfield
  {title} {\emph {\enquote {\bibinfo {title} {{Lee-Wick Black Holes}},}\
  }}\href@noop {} {\  (\bibinfo {year} {2016}{\natexlab{c}})},\ \Eprint
  {http://arxiv.org/abs/1611.03650} {arXiv:1611.03650 [gr-qc]} \BibitemShut
  {NoStop}%
\bibitem [{\citenamefont {Bambi}\ \emph
  {et~al.}(2016{\natexlab{d}})\citenamefont {Bambi}, \citenamefont {Modesto},\
  and\ \citenamefont {Rachwal}}]{Bambi:2016wdn}%
  \BibitemOpen
  \bibfield  {author} {\bibinfo {author} {\bibfnamefont {C.}~\bibnamefont
  {Bambi}}, \bibinfo {author} {\bibfnamefont {L.}~\bibnamefont {Modesto}}, \
  and\ \bibinfo {author} {\bibfnamefont {L.}~\bibnamefont {Rachwal}},\
  }\bibfield  {title} {\emph {\enquote {\bibinfo {title} {{Spacetime
  completeness of non-singular black holes in conformal gravity}},}\
  }}\href@noop {} {\  (\bibinfo {year} {2016}{\natexlab{d}})},\ \Eprint
  {http://arxiv.org/abs/1611.00865} {arXiv:1611.00865 [gr-qc]} \BibitemShut
  {NoStop}%
\bibitem [{\citenamefont {Frolov}(2016)}]{Frolov:2016pav}%
  \BibitemOpen
  \bibfield  {author} {\bibinfo {author} {\bibfnamefont {V.~P.}\ \bibnamefont
  {Frolov}},\ }\bibfield  {title} {\emph {\enquote {\bibinfo {title} {{Notes on
  non-singular models of black holes}},}\ }}\href {\doibase
  10.1103/PhysRevD.94.104056} {\bibfield  {journal} {\bibinfo  {journal} {Phys.
  Rev.}\ }\textbf {\bibinfo {volume} {D94}},\ \bibinfo {pages} {104056}
  (\bibinfo {year} {2016})},\ \Eprint {http://arxiv.org/abs/1609.01758}
  {arXiv:1609.01758 [gr-qc]} \BibitemShut {NoStop}%
\bibitem [{\citenamefont {Markov}(1982)}]{Markov:1982}%
  \BibitemOpen
  \bibfield  {author} {\bibinfo {author} {\bibfnamefont {M.}~\bibnamefont
  {Markov}},\ }\bibfield  {title} {\emph {\enquote {\bibinfo {title} {{Limiting
  density of matter as a universal law of nature}},}\ }}\href@noop {}
  {\bibfield  {journal} {\bibinfo  {journal} {JETP Letters}\ }\textbf {\bibinfo
  {volume} {36}},\ \bibinfo {pages} {266} (\bibinfo {year} {1982})}\BibitemShut
  {NoStop}%
\bibitem [{\citenamefont {Markov}(1984)}]{Markov:1984ii}%
  \BibitemOpen
  \bibfield  {author} {\bibinfo {author} {\bibfnamefont {M.}~\bibnamefont
  {Markov}},\ }\bibfield  {title} {\emph {\enquote {\bibinfo {title} {{Problems
  of a Perpetually Oscillating Universe}},}\ }}\href {\doibase
  10.1016/0003-4916(84)90004-6} {\bibfield  {journal} {\bibinfo  {journal}
  {Annals Phys.}\ }\textbf {\bibinfo {volume} {155}},\ \bibinfo {pages} {333}
  (\bibinfo {year} {1984})}\BibitemShut {NoStop}%
\bibitem [{\citenamefont {Polchinski}(1989)}]{Polchinski:1989ae}%
  \BibitemOpen
  \bibfield  {author} {\bibinfo {author} {\bibfnamefont {J.}~\bibnamefont
  {Polchinski}},\ }\bibfield  {title} {\emph {\enquote {\bibinfo {title}
  {{Decoupling Versus Excluded Volume or Return of the Giant Wormholes}},}\
  }}\href {\doibase 10.1016/0550-3213(89)90499-9} {\bibfield  {journal}
  {\bibinfo  {journal} {Nucl.Phys.}\ }\textbf {\bibinfo {volume} {B325}},\
  \bibinfo {pages} {619} (\bibinfo {year} {1989})}\BibitemShut {NoStop}%
\bibitem [{\citenamefont {Frolov}\ and\ \citenamefont
  {Vilkovisky}(1981)}]{Frolov:1981mz}%
  \BibitemOpen
  \bibfield  {author} {\bibinfo {author} {\bibfnamefont {V.~P.}\ \bibnamefont
  {Frolov}}\ and\ \bibinfo {author} {\bibfnamefont {G.}~\bibnamefont
  {Vilkovisky}},\ }\bibfield  {title} {\emph {\enquote {\bibinfo {title}
  {{Spherically Symmetric Collapse in Quantum Gravity}},}\ }}\href {\doibase
  10.1016/0370-2693(81)90542-6} {\bibfield  {journal} {\bibinfo  {journal}
  {Phys.Lett.}\ }\textbf {\bibinfo {volume} {B106}},\ \bibinfo {pages} {307}
  (\bibinfo {year} {1981})}\BibitemShut {NoStop}%
\bibitem [{\citenamefont {Roman}\ and\ \citenamefont
  {Bergmann}(1983)}]{Roman:1983zza}%
  \BibitemOpen
  \bibfield  {author} {\bibinfo {author} {\bibfnamefont {T.~A.}\ \bibnamefont
  {Roman}}\ and\ \bibinfo {author} {\bibfnamefont {P.~G.}\ \bibnamefont
  {Bergmann}},\ }\bibfield  {title} {\emph {\enquote {\bibinfo {title}
  {{Stellar collapse without singularities?}}}\ }}\href {\doibase
  10.1103/PhysRevD.28.1265} {\bibfield  {journal} {\bibinfo  {journal} {Phys.
  Rev.}\ }\textbf {\bibinfo {volume} {D28}},\ \bibinfo {pages} {1265} (\bibinfo
  {year} {1983})}\BibitemShut {NoStop}%
\bibitem [{\citenamefont {Borde}(1997)}]{Borde:1996df}%
  \BibitemOpen
  \bibfield  {author} {\bibinfo {author} {\bibfnamefont {A.}~\bibnamefont
  {Borde}},\ }\bibfield  {title} {\emph {\enquote {\bibinfo {title} {{Regular
  black holes and topology change}},}\ }}\href {\doibase
  10.1103/PhysRevD.55.7615} {\bibfield  {journal} {\bibinfo  {journal} {Phys.
  Rev.}\ }\textbf {\bibinfo {volume} {D55}},\ \bibinfo {pages} {7615} (\bibinfo
  {year} {1997})},\ \Eprint {http://arxiv.org/abs/gr-qc/9612057}
  {arXiv:gr-qc/9612057 [gr-qc]} \BibitemShut {NoStop}%
\bibitem [{\citenamefont {Hayward}(2006)}]{Hayward:2005gi}%
  \BibitemOpen
  \bibfield  {author} {\bibinfo {author} {\bibfnamefont {S.~A.}\ \bibnamefont
  {Hayward}},\ }\bibfield  {title} {\emph {\enquote {\bibinfo {title}
  {{Formation and evaporation of regular black holes}},}\ }}\href {\doibase
  10.1103/PhysRevLett.96.031103} {\bibfield  {journal} {\bibinfo  {journal}
  {Phys.Rev.Lett.}\ }\textbf {\bibinfo {volume} {96}},\ \bibinfo {pages}
  {031103} (\bibinfo {year} {2006})},\ \Eprint
  {http://arxiv.org/abs/gr-qc/0506126} {arXiv:gr-qc/0506126 [gr-qc]}
  \BibitemShut {NoStop}%
\bibitem [{\citenamefont {Bambi}\ \emph {et~al.}(2014)\citenamefont {Bambi},
  \citenamefont {Malafarina},\ and\ \citenamefont {Modesto}}]{Bambi:2013gva}%
  \BibitemOpen
  \bibfield  {author} {\bibinfo {author} {\bibfnamefont {C.}~\bibnamefont
  {Bambi}}, \bibinfo {author} {\bibfnamefont {D.}~\bibnamefont {Malafarina}}, \
  and\ \bibinfo {author} {\bibfnamefont {L.}~\bibnamefont {Modesto}},\
  }\bibfield  {title} {\emph {\enquote {\bibinfo {title} {{Terminating black
  holes in asymptotically free quantum gravity}},}\ }}\href {\doibase
  10.1140/epjc/s10052-014-2767-9} {\bibfield  {journal} {\bibinfo  {journal}
  {Eur. Phys. J.}\ }\textbf {\bibinfo {volume} {C74}},\ \bibinfo {pages} {2767}
  (\bibinfo {year} {2014})},\ \Eprint {http://arxiv.org/abs/1306.1668}
  {arXiv:1306.1668 [gr-qc]} \BibitemShut {NoStop}%
\bibitem [{\citenamefont {Bambi}\ \emph {et~al.}(2013)\citenamefont {Bambi},
  \citenamefont {Malafarina},\ and\ \citenamefont {Modesto}}]{Bambi:2013caa}%
  \BibitemOpen
  \bibfield  {author} {\bibinfo {author} {\bibfnamefont {C.}~\bibnamefont
  {Bambi}}, \bibinfo {author} {\bibfnamefont {D.}~\bibnamefont {Malafarina}}, \
  and\ \bibinfo {author} {\bibfnamefont {L.}~\bibnamefont {Modesto}},\
  }\bibfield  {title} {\emph {\enquote {\bibinfo {title} {{Non-singular
  quantum-inspired gravitational collapse}},}\ }}\href {\doibase
  10.1103/PhysRevD.88.044009} {\bibfield  {journal} {\bibinfo  {journal} {Phys.
  Rev.}\ }\textbf {\bibinfo {volume} {D88}},\ \bibinfo {pages} {044009}
  (\bibinfo {year} {2013})},\ \Eprint {http://arxiv.org/abs/1305.4790}
  {arXiv:1305.4790 [gr-qc]} \BibitemShut {NoStop}%
\bibitem [{\citenamefont {Hawking}(2014)}]{Hawking:2014tga}%
  \BibitemOpen
  \bibfield  {author} {\bibinfo {author} {\bibfnamefont {S.~W.}\ \bibnamefont
  {Hawking}},\ }\bibfield  {title} {\emph {\enquote {\bibinfo {title}
  {{Information Preservation and Weather Forecasting for Black Holes}},}\
  }}\href@noop {} {\  (\bibinfo {year} {2014})},\ \Eprint
  {http://arxiv.org/abs/1401.5761} {arXiv:1401.5761 [hep-th]} \BibitemShut
  {NoStop}%
\bibitem [{\citenamefont {Frolov}(2014)}]{Frolov:2014jva}%
  \BibitemOpen
  \bibfield  {author} {\bibinfo {author} {\bibfnamefont {V.~P.}\ \bibnamefont
  {Frolov}},\ }\bibfield  {title} {\emph {\enquote {\bibinfo {title}
  {{Information loss problem and a 'black hole` model with a closed apparent
  horizon}},}\ }}\href {\doibase 10.1007/JHEP05(2014)049} {\bibfield  {journal}
  {\bibinfo  {journal} {JHEP}\ }\textbf {\bibinfo {volume} {1405}},\ \bibinfo
  {pages} {049} (\bibinfo {year} {2014})},\ \Eprint
  {http://arxiv.org/abs/1402.5446} {arXiv:1402.5446 [hep-th]} \BibitemShut
  {NoStop}%
\bibitem [{\citenamefont {Bardeen}(2014)}]{Bardeen:2014uaa}%
  \BibitemOpen
  \bibfield  {author} {\bibinfo {author} {\bibfnamefont {J.~M.}\ \bibnamefont
  {Bardeen}},\ }\bibfield  {title} {\emph {\enquote {\bibinfo {title} {{Black
  hole evaporation without an event horizon}},}\ }}\href@noop {} {\  (\bibinfo
  {year} {2014})},\ \Eprint {http://arxiv.org/abs/1406.4098} {arXiv:1406.4098
  [gr-qc]} \BibitemShut {NoStop}%
\bibitem [{\citenamefont {Haggard}\ and\ \citenamefont
  {Rovelli}(2015{\natexlab{a}})}]{Haggard:2014rza}%
  \BibitemOpen
  \bibfield  {author} {\bibinfo {author} {\bibfnamefont {H.~M.}\ \bibnamefont
  {Haggard}}\ and\ \bibinfo {author} {\bibfnamefont {C.}~\bibnamefont
  {Rovelli}},\ }\bibfield  {title} {\emph {\enquote {\bibinfo {title}
  {{Quantum-gravity effects outside the horizon spark black to white hole
  tunneling}},}\ }}\href {\doibase 10.1103/PhysRevD.92.104020} {\bibfield
  {journal} {\bibinfo  {journal} {Phys. Rev.}\ }\textbf {\bibinfo {volume}
  {D92}},\ \bibinfo {pages} {104020} (\bibinfo {year} {2015}{\natexlab{a}})},\
  \Eprint {http://arxiv.org/abs/1407.0989} {arXiv:1407.0989 [gr-qc]}
  \BibitemShut {NoStop}%
\bibitem [{\citenamefont {Barrau}\ \emph {et~al.}(2016)\citenamefont {Barrau},
  \citenamefont {Bolliet}, \citenamefont {Vidotto},\ and\ \citenamefont
  {Weimer}}]{Barrau:2015uca}%
  \BibitemOpen
  \bibfield  {author} {\bibinfo {author} {\bibfnamefont {A.}~\bibnamefont
  {Barrau}}, \bibinfo {author} {\bibfnamefont {B.}~\bibnamefont {Bolliet}},
  \bibinfo {author} {\bibfnamefont {F.}~\bibnamefont {Vidotto}}, \ and\
  \bibinfo {author} {\bibfnamefont {C.}~\bibnamefont {Weimer}},\ }\bibfield
  {title} {\emph {\enquote {\bibinfo {title} {{Phenomenology of bouncing black
  holes in quantum gravity: a closer look}},}\ }}\href {\doibase
  10.1088/1475-7516/2016/02/022} {\bibfield  {journal} {\bibinfo  {journal}
  {JCAP}\ }\textbf {\bibinfo {volume} {1602}},\ \bibinfo {pages} {022}
  (\bibinfo {year} {2016})},\ \Eprint {http://arxiv.org/abs/1507.05424}
  {arXiv:1507.05424 [gr-qc]} \BibitemShut {NoStop}%
\bibitem [{\citenamefont {Haggard}\ and\ \citenamefont
  {Rovelli}(2015{\natexlab{b}})}]{Haggard:2015iya}%
  \BibitemOpen
  \bibfield  {author} {\bibinfo {author} {\bibfnamefont {H.~M.}\ \bibnamefont
  {Haggard}}\ and\ \bibinfo {author} {\bibfnamefont {C.}~\bibnamefont
  {Rovelli}},\ }\bibfield  {title} {\emph {\enquote {\bibinfo {title} {{Black
  to white hole tunneling: An exact classical solution}},}\ }}\href {\doibase
  10.1142/S0217751X15450153} {\bibfield  {journal} {\bibinfo  {journal} {Int.
  J. Mod. Phys.}\ }\textbf {\bibinfo {volume} {A30}},\ \bibinfo {pages}
  {1545015} (\bibinfo {year} {2015}{\natexlab{b}})}\BibitemShut {NoStop}%
\bibitem [{\citenamefont {Bolashenko}\ and\ \citenamefont
  {Frolov}(1986)}]{Bolashenko:1986mr}%
  \BibitemOpen
  \bibfield  {author} {\bibinfo {author} {\bibfnamefont {P.~A.}\ \bibnamefont
  {Bolashenko}}\ and\ \bibinfo {author} {\bibfnamefont {V.~P.}\ \bibnamefont
  {Frolov}},\ }in\ \href@noop {} {\emph {\bibinfo {booktitle} {{Physical
  Effects in the Gravitational Fields of Black Holes , Proceedings of the
  Lebedev Physics Institute of the Academy of Sciences of the USSR, v.169}}}},\
  \bibinfo {editor} {edited by\ \bibinfo {editor} {\bibfnamefont {M.~A.}\
  \bibnamefont {Markov}}}\ (\bibinfo  {publisher} {Commack, N.Y., Nova
  Science},\ \bibinfo {year} {1986})\ pp.\ \bibinfo {pages}
  {159--168}\BibitemShut {NoStop}%
\bibitem [{\citenamefont {Bianchi}\ \emph {et~al.}(2015)\citenamefont
  {Bianchi}, \citenamefont {De~Lorenzo},\ and\ \citenamefont
  {Smerlak}}]{Bianchi:2014bma}%
  \BibitemOpen
  \bibfield  {author} {\bibinfo {author} {\bibfnamefont {E.}~\bibnamefont
  {Bianchi}}, \bibinfo {author} {\bibfnamefont {T.}~\bibnamefont {De~Lorenzo}},
  \ and\ \bibinfo {author} {\bibfnamefont {M.}~\bibnamefont {Smerlak}},\
  }\bibfield  {title} {\emph {\enquote {\bibinfo {title} {{Entanglement entropy
  production in gravitational collapse: covariant regularization and solvable
  models}},}\ }}\href {\doibase 10.1007/JHEP06(2015)180} {\bibfield  {journal}
  {\bibinfo  {journal} {JHEP}\ }\textbf {\bibinfo {volume} {06}},\ \bibinfo
  {pages} {180} (\bibinfo {year} {2015})},\ \Eprint
  {http://arxiv.org/abs/1409.0144} {arXiv:1409.0144 [hep-th]} \BibitemShut
  {NoStop}%
\bibitem [{\citenamefont {Christensen}\ and\ \citenamefont
  {Fulling}(1977)}]{Christensen:1977jc}%
  \BibitemOpen
  \bibfield  {author} {\bibinfo {author} {\bibfnamefont {S.~M.}\ \bibnamefont
  {Christensen}}\ and\ \bibinfo {author} {\bibfnamefont {S.~A.}\ \bibnamefont
  {Fulling}},\ }\bibfield  {title} {\emph {\enquote {\bibinfo {title} {{Trace
  Anomalies and the Hawking Effect}},}\ }}\href {\doibase
  10.1103/PhysRevD.15.2088} {\bibfield  {journal} {\bibinfo  {journal} {Phys.
  Rev.}\ }\textbf {\bibinfo {volume} {D15}},\ \bibinfo {pages} {2088} (\bibinfo
  {year} {1977})}\BibitemShut {NoStop}%
\bibitem [{\citenamefont {Frolov}\ and\ \citenamefont
  {Novikov}(1998)}]{Frolov:1998wf}%
  \BibitemOpen
  \bibfield  {author} {\bibinfo {author} {\bibfnamefont {V.}~\bibnamefont
  {Frolov}}\ and\ \bibinfo {author} {\bibfnamefont {I.}~\bibnamefont
  {Novikov}},\ }\href@noop {} {\emph {\bibinfo {title} {{Black hole physics:
  Basic concepts and new developments}}}}\ (\bibinfo  {publisher} {Kluwer Acad.
  Publ.},\ \bibinfo {year} {1998})\BibitemShut {NoStop}%
\bibitem [{Note1()}]{Note1}%
  \BibitemOpen
  \bibinfo {note} {Certainly, the question of how good is this approximation
  for the calculation of the energy flux from the black hole interior, should
  be studied.}\BibitemShut {Stop}%
\bibitem [{\citenamefont {Frolov}\ and\ \citenamefont
  {Vilkovisky}(1984)}]{Frolov:1984}%
  \BibitemOpen
  \bibfield  {author} {\bibinfo {author} {\bibfnamefont {V.~P.}\ \bibnamefont
  {Frolov}}\ and\ \bibinfo {author} {\bibfnamefont {G.~A.}\ \bibnamefont
  {Vilkovisky}},\ }in\ \href@noop {} {\emph {\bibinfo {booktitle} {{Quantum
  Gravity}}}}\ (\bibinfo  {publisher} {Springer US},\ \bibinfo {year} {1984})\
  pp.\ \bibinfo {pages} {267--290}\BibitemShut {NoStop}%
\bibitem [{\citenamefont {Vilkovisky}(1985)}]{Vilkovisky:1985}%
  \BibitemOpen
  \bibfield  {author} {\bibinfo {author} {\bibfnamefont {G.~A.}\ \bibnamefont
  {Vilkovisky}},\ }in\ \href@noop {} {\emph {\bibinfo {booktitle} {{Quantum
  theory of gravity}}}}\ (\bibinfo  {publisher} {Adam Hilger Limited, Bristol
  UK},\ \bibinfo {year} {1985})\ pp.\ \bibinfo {pages} {169--209}\BibitemShut
  {NoStop}%
\bibitem [{\citenamefont {De~Lorenzo}(2014)}]{Lorenzo}%
  \BibitemOpen
  \bibfield  {author} {\bibinfo {author} {\bibfnamefont {T.}~\bibnamefont
  {De~Lorenzo}},\ }\emph {\bibinfo {title} {{Investigating static and dynamic
  non-singular black hole}}},\ \href
  {https://etd.adm.unipi.it/theses/available/etd-11232014-215047/unrestricted/Master_thesis_DeLorenzo.pdf}
  {Master's thesis},\ \bibinfo  {school} {University of Pisa} (\bibinfo {year}
  {2014})\BibitemShut {NoStop}%
\bibitem [{Note2()}]{Note2}%
  \BibitemOpen
  \bibinfo {note} {We use an ambiguity in the choice of $\lambda $ in such a
  way, that $\protect \mathaccentV {dot}05F{x}^{\mu }$ coincides with the
  four-momentum of the photon.}\BibitemShut {Stop}%
\bibitem [{Note3()}]{Note3}%
  \BibitemOpen
  \bibinfo {note} {Similar sandwich type black holes were discussed earlier in
  the interesting paper \cite {Bianchi:2014bma}, where the problem of the
  entanglement entropy of evaporating black holes was considered. Namely, the
  authors of this paper assumed that the metric between two null collapsing
  shell is constructed by gluing together of the de Sitter metric (for small
  radius) and the Schwarzschild one (for large radius). They argued that such a
  metric can be used as an approximation for the Hayward spacetime. In the
  first part of our paper we discuss a sandwich black hole with exact Hayward
  interior. What is more important, we discuss in the second part of the paper
  metrics between the shells which have an additional new property: a
  non-trivial redshift factor, which was ignored in the paper \cite
  {Bianchi:2014bma}.}\BibitemShut {Stop}%
\bibitem [{Note4()}]{Note4}%
  \BibitemOpen
  \bibinfo {note} {Sometimes one can see in the literature a statement that all
  the particles of the Hawking radiation are created during the formation of
  the black hole. The relation (\ref {DEE}) shows that such a statement is
  totally wrong. A decision when to switch-off the black hole can be made any
  time after its formation, so that $\Delta V$ is an arbitrary
  parameter.}\BibitemShut {Stop}%
\bibitem [{\citenamefont {Barrabes}\ \emph {et~al.}(1999)\citenamefont
  {Barrabes}, \citenamefont {Frolov},\ and\ \citenamefont
  {Parentani}}]{Barrabes:1998iw}%
  \BibitemOpen
  \bibfield  {author} {\bibinfo {author} {\bibfnamefont {C.}~\bibnamefont
  {Barrabes}}, \bibinfo {author} {\bibfnamefont {V.~P.}\ \bibnamefont
  {Frolov}}, \ and\ \bibinfo {author} {\bibfnamefont {R.}~\bibnamefont
  {Parentani}},\ }\bibfield  {title} {\emph {\enquote {\bibinfo {title}
  {{Metric fluctuation corrections to Hawking radiation}},}\ }}\href {\doibase
  10.1103/PhysRevD.59.124010} {\bibfield  {journal} {\bibinfo  {journal} {Phys.
  Rev.}\ }\textbf {\bibinfo {volume} {D59}},\ \bibinfo {pages} {124010}
  (\bibinfo {year} {1999})},\ \Eprint {http://arxiv.org/abs/gr-qc/9812076}
  {arXiv:gr-qc/9812076 [gr-qc]} \BibitemShut {NoStop}%
\bibitem [{\citenamefont {Barrabes}\ \emph {et~al.}(2000)\citenamefont
  {Barrabes}, \citenamefont {Frolov},\ and\ \citenamefont
  {Parentani}}]{Barrabes:2000fr}%
  \BibitemOpen
  \bibfield  {author} {\bibinfo {author} {\bibfnamefont {C.}~\bibnamefont
  {Barrabes}}, \bibinfo {author} {\bibfnamefont {V.~P.}\ \bibnamefont
  {Frolov}}, \ and\ \bibinfo {author} {\bibfnamefont {R.}~\bibnamefont
  {Parentani}},\ }\bibfield  {title} {\emph {\enquote {\bibinfo {title}
  {{Stochastically fluctuating black hole geometry, Hawking radiation and the
  transPlanckian problem}},}\ }}\href {\doibase 10.1103/PhysRevD.62.044020}
  {\bibfield  {journal} {\bibinfo  {journal} {Phys. Rev.}\ }\textbf {\bibinfo
  {volume} {D62}},\ \bibinfo {pages} {044020} (\bibinfo {year} {2000})},\
  \Eprint {http://arxiv.org/abs/gr-qc/0001102} {arXiv:gr-qc/0001102 [gr-qc]}
  \BibitemShut {NoStop}%
\bibitem [{\citenamefont {De~Lorenzo}\ \emph {et~al.}(2015)\citenamefont
  {De~Lorenzo}, \citenamefont {Pacilio}, \citenamefont {Rovelli},\ and\
  \citenamefont {Speziale}}]{DeLorenzo:2014pta}%
  \BibitemOpen
  \bibfield  {author} {\bibinfo {author} {\bibfnamefont {T.}~\bibnamefont
  {De~Lorenzo}}, \bibinfo {author} {\bibfnamefont {C.}~\bibnamefont {Pacilio}},
  \bibinfo {author} {\bibfnamefont {C.}~\bibnamefont {Rovelli}}, \ and\
  \bibinfo {author} {\bibfnamefont {S.}~\bibnamefont {Speziale}},\ }\bibfield
  {title} {\emph {\enquote {\bibinfo {title} {{On the Effective Metric of a
  Planck Star}},}\ }}\href {\doibase 10.1007/s10714-015-1882-8} {\bibfield
  {journal} {\bibinfo  {journal} {Gen. Rel. Grav.}\ }\textbf {\bibinfo {volume}
  {47}},\ \bibinfo {pages} {41} (\bibinfo {year} {2015})},\ \Eprint
  {http://arxiv.org/abs/1412.6015} {arXiv:1412.6015 [gr-qc]} \BibitemShut
  {NoStop}%
\bibitem [{\citenamefont {De~Lorenzo}\ \emph {et~al.}(2016)\citenamefont
  {De~Lorenzo}, \citenamefont {Giusti},\ and\ \citenamefont
  {Speziale}}]{DeLorenzo:2015taa}%
  \BibitemOpen
  \bibfield  {author} {\bibinfo {author} {\bibfnamefont {T.}~\bibnamefont
  {De~Lorenzo}}, \bibinfo {author} {\bibfnamefont {A.}~\bibnamefont {Giusti}},
  \ and\ \bibinfo {author} {\bibfnamefont {S.}~\bibnamefont {Speziale}},\
  }\bibfield  {title} {\emph {\enquote {\bibinfo {title} {{Non-singular
  rotating black hole with a time delay in the center}},}\ }}\href {\doibase
  10.1007/s10714-016-2026-5, 10.1007/s10714-016-2105-7} {\bibfield  {journal}
  {\bibinfo  {journal} {Gen. Rel. Grav.}\ }\textbf {\bibinfo {volume} {48}},\
  \bibinfo {pages} {31} (\bibinfo {year} {2016})},\ \bibinfo {note} {[Erratum:
  Gen. Rel. Grav.48,no.8,111(2016)]},\ \Eprint
  {http://arxiv.org/abs/1510.08828} {arXiv:1510.08828 [gr-qc]} \BibitemShut
  {NoStop}%
\bibitem [{\citenamefont {Trodden}\ \emph {et~al.}(1993)\citenamefont
  {Trodden}, \citenamefont {Mukhanov},\ and\ \citenamefont
  {Brandenberger}}]{Trodden:1993dm}%
  \BibitemOpen
  \bibfield  {author} {\bibinfo {author} {\bibfnamefont {M.}~\bibnamefont
  {Trodden}}, \bibinfo {author} {\bibfnamefont {V.~F.}\ \bibnamefont
  {Mukhanov}}, \ and\ \bibinfo {author} {\bibfnamefont {R.~H.}\ \bibnamefont
  {Brandenberger}},\ }\bibfield  {title} {\emph {\enquote {\bibinfo {title} {{A
  Nonsingular two-dimensional black hole}},}\ }}\href {\doibase
  10.1016/0370-2693(93)91032-I} {\bibfield  {journal} {\bibinfo  {journal}
  {Phys. Lett.}\ }\textbf {\bibinfo {volume} {B316}},\ \bibinfo {pages} {483}
  (\bibinfo {year} {1993})},\ \Eprint {http://arxiv.org/abs/hep-th/9305111}
  {arXiv:hep-th/9305111 [hep-th]} \BibitemShut {NoStop}%
\bibitem [{\citenamefont {Easson}(2003)}]{Easson:2002tg}%
  \BibitemOpen
  \bibfield  {author} {\bibinfo {author} {\bibfnamefont {D.~A.}\ \bibnamefont
  {Easson}},\ }\bibfield  {title} {\emph {\enquote {\bibinfo {title} {{Hawking
  radiation of nonsingular black holes in two-dimensions}},}\ }}\href {\doibase
  10.1088/1126-6708/2003/02/037} {\bibfield  {journal} {\bibinfo  {journal}
  {JHEP}\ }\textbf {\bibinfo {volume} {02}},\ \bibinfo {pages} {037} (\bibinfo
  {year} {2003})},\ \Eprint {http://arxiv.org/abs/hep-th/0210016}
  {arXiv:hep-th/0210016 [hep-th]} \BibitemShut {NoStop}%
\bibitem [{\citenamefont {Taves}\ and\ \citenamefont
  {Kunstatter}(2014)}]{Taves:2014laa}%
  \BibitemOpen
  \bibfield  {author} {\bibinfo {author} {\bibfnamefont {T.}~\bibnamefont
  {Taves}}\ and\ \bibinfo {author} {\bibfnamefont {G.}~\bibnamefont
  {Kunstatter}},\ }\bibfield  {title} {\emph {\enquote {\bibinfo {title}
  {{Modelling the evaporation of nonsingular black holes}},}\ }}\href {\doibase
  10.1103/PhysRevD.90.124062} {\bibfield  {journal} {\bibinfo  {journal} {Phys.
  Rev.}\ }\textbf {\bibinfo {volume} {D90}},\ \bibinfo {pages} {124062}
  (\bibinfo {year} {2014})},\ \Eprint {http://arxiv.org/abs/1408.1444}
  {arXiv:1408.1444 [gr-qc]} \BibitemShut {NoStop}%
\bibitem [{\citenamefont {Kunstatter}\ \emph {et~al.}(2016)\citenamefont
  {Kunstatter}, \citenamefont {Maeda},\ and\ \citenamefont
  {Taves}}]{Kunstatter:2015vxa}%
  \BibitemOpen
  \bibfield  {author} {\bibinfo {author} {\bibfnamefont {G.}~\bibnamefont
  {Kunstatter}}, \bibinfo {author} {\bibfnamefont {H.}~\bibnamefont {Maeda}}, \
  and\ \bibinfo {author} {\bibfnamefont {T.}~\bibnamefont {Taves}},\ }\bibfield
   {title} {\emph {\enquote {\bibinfo {title} {{New 2D dilaton gravity for
  nonsingular black holes}},}\ }}\href {\doibase
  10.1088/0264-9381/33/10/105005} {\bibfield  {journal} {\bibinfo  {journal}
  {Class. Quant. Grav.}\ }\textbf {\bibinfo {volume} {33}},\ \bibinfo {pages}
  {105005} (\bibinfo {year} {2016})},\ \Eprint
  {http://arxiv.org/abs/1509.06746} {arXiv:1509.06746 [gr-qc]} \BibitemShut
  {NoStop}%
\end{thebibliography}
\end{document}